\documentclass[amsmath,amssymb,aps,prd,showkeys,reprint,%
floatfix,groupedaddress,superscriptaddress,amsmath,amssymb]{revtex4-2}

\usepackage[breaklinks = true, colorlinks = true,%
 linkcolor = blue,citecolor = blue]{hyperref}
\usepackage{graphicx}
\usepackage{xcolor}
\usepackage{MnSymbol}

\newcommand{\bB}{\boldsymbol{B}}
\newcommand{\bD}{\boldsymbol{D}}
\newcommand{\bE}{\boldsymbol{E}}
\newcommand{\bgamma}{\boldsymbol{\gamma}}
\newcommand{\bnabla}{\boldsymbol{\nabla}}
\newcommand{\bomega}{\boldsymbol{\omega}}
\newcommand{\bSigma}{\boldsymbol{\Sigma}}
\newcommand{\be}{\boldsymbol{e}}
\newcommand{\bx}{\boldsymbol{x}}
\newcommand{\bl}{\boldsymbol{l}}
\newcommand{\bs}{\boldsymbol{s}}
\newcommand{\bj}{\boldsymbol{j}}
\newcommand{\bJ}{\boldsymbol{J}}

\newcommand{\avej}{\mathfrak{j}}
\newcommand{\aver}{\mathfrak{r}}

\begin{document}

\title{Mode Decomposed Chiral Magnetic Effect and Rotating Fermions}

\author{Kenji Fukushima}
\email{fuku@nt.phys.s.u-tokyo.ac.jp}
\affiliation{Department of Physics, The University of Tokyo,
  7-3-1 Hongo, Bunkyo-ku, Tokyo 113-0033, Japan}

\author{Takuya Shimazaki}
\email{shimazaki@nt.phys.s.u-tokyo.ac.jp}
\affiliation{Department of Physics, The University of Tokyo,
  7-3-1 Hongo, Bunkyo-ku, Tokyo 113-0033, Japan}

\author{Lingxiao Wang}
\email{wlx15@tsinghua.org.cn}
\affiliation{Physics Department, Tsinghua University and Collaborative
  Innovation Center of Quantum Matter, Beijing 100084, China}
\affiliation{Department of Physics, The University of Tokyo,
  7-3-1 Hongo, Bunkyo-ku, Tokyo 113-0033, Japan}

\begin{abstract}
  We present a novel perspective to characterize the chiral magnetic
  and related effects in terms of angular decomposed modes.  We find
  that the vector current and the chirality density are connected
  through a surprisingly simple relation for all the modes and any
  mass, which defines the mode decomposed chiral magnetic effect in
  such a way free from the chiral chemical potential.  The mode
  decomposed formulation is useful also to investigate properties of
  rotating fermions.  For demonstration we give an intuitive account
  for a nonzero density emerging from a combination of rotation and
  magnetic field as well as an approach to the chiral vortical effect
  at finite density.
\end{abstract}
\maketitle

\section{Introduction}

Chiral fermions exhibit fascinating transport properties, the origin
of which is traced back to the quantum anomaly associated with chiral
symmetry.  Gauge theories including quantum chromodynamics (QCD) may
accommodate topologically winding configurations and a chirality
imbalance is induced~\cite{Adler:1969gk,Bell:1969ts}.  Thanks to the
index theorem, one can quantify the chirality change in response to
the winding number even without solving QCD problems.  There are some
indirect evidences for the chiral anomaly such as the decay rate of
$\pi^0\to2\gamma$, anomalously heavy $\eta'$
meson~\cite{Witten:1979vv,Veneziano:1979ec}, etc.  It is still an
ambitious problem to establish more direct experimental probes to the
chiral anomaly~\cite{Kharzeev:1998kz}.

The chiral magnetic effect (CME) is a well investigated example of
macroscopic realization of the chiral anomaly, which was originally
proposed in the context of the relativistic heavy-ion collision
experiment~\cite{Kharzeev:2007jp} (see also
Ref.~\cite{Vilenkin:1980fu} for an earlier work) and an elegant
formula with the chiral chemical potential $\mu_5$ was found in
Ref.~\cite{Fukushima:2008xe}.  The CME generates a vector current
$\bj$ in parallel to an external magnetic field $\bB$ in the presence
of a finite chirality imbalance.  In ordinary matter such a vector
current proportional to $\bB$ is prohibited by parity and
time-reversal symmetry, so the CME requires the chirality imbalance.
In other words the coefficient in front of $\bB$ in the CME, i.e., the
chiral magnetic conductivity~\cite{Kharzeev:2009pj}, must be parity
odd and time-reversal even.  Interestingly, the CME is anomaly
protected and the chiral magnetic conductivity at zero frequency is
unaffected by higher-order corrections.  There are a number of
theoretical and experimental efforts to detect CME signatures;
parity-odd fluctuations (called the $\gamma$-correlator) in the
heavy-ion
collisions~\cite{Abelev:2009ac,Abelev:2009ad,Kharzeev:2010gr,Abelev:2012pa,Huang:2015oca,Kharzeev:2015znc}
and the negative magnetoresistance in Weyl/Dirac
semimetals~\cite{Son:2012bg,Li:2014bha,Huang:2015eia,Li:2016vlc} are
representative examples.

A finite chirality imbalance plays a pivotal role in the CME, which
supposedly results from topological excitations in the QCD case.  For
theoretical convenience a finite chirality is imposed conventionally
by $\mu_5$ coupled to the chirality charge.  Sometimes, however,
$\mu_5$ has caused controversies -- the chemical potential is
generally well-defined for a conserved charge but the chirality charge
is not conserved due to the chiral anomaly.  Since it is not a conserved
charge, $\mu_5$ must be vanishing in equilibrium.  In other words a
finite $\mu_5$ as a theoretical device makes sense only out of
equilibrium.  It would be therefore desirable to setup the CME without
$\mu_5$.  In fact a background electromagnetic field like $\bE\propto
\bB$ is an alternative of $\mu_5$ used for experiments in Weyl/Dirac
semimetals.  For more details on the absence of the CME in
equilibrium, see Ref.~\cite{Copinger:2018ftr}.

Let us turn to other intriguing phenomena in a chiral medium similar
to the CME, namely, the chiral separation effect
(CSE)~\cite{Son:2004tq,Metlitski:2005pr,Newman:2005as} and the chiral
vortical effect
(CVE)~\cite{Vilenkin:1979ui,Vilenkin:1980zv,Erdmenger:2008rm,Son:2009tf,Landsteiner:2011cp,Flachi:2017vlp,Abramchuk:2018jhd}.
In later discussions we will shed light on the CSE and the CVE as well
as the CME using the mode decomposed wave-functions.  In the CSE an
axial-vector current $\bj_5$ is generated in parallel to an external
magnetic field $\bB$ in the presence of the chemical potential $\mu$.
The CVE is quite analogous to the CSE;  an axial-vector current is
induced by a vorticity $\bomega$ instead of
$\bB$~\cite{Kharzeev:2007tn,Stephanov:2012ki,McInnes:2016dwk}.
Contrary to the CME, the CSE and the CVE are not anomaly-protected and
their coefficients could be affected by infrared scales such as the
mass, the temperature, and so
on~\cite{Gorbar:2013upa,Guo:2016dnm,Flachi:2017vlp,Lin:2018aon,Wang:2019moi}.
Such a clear distinction between the CME and the CSE/CVE will be
unraveled in our mode decomposed analysis.  Our calculations will lead
to a surprisingly simple relation between the vector current and the
chirality density, which had been unseen until we made a mode
decomposed formulation.

Interestingly, both strong magnetic field $\bB$ and large vorticity
$\bomega$ are highly relevant to the heavy-ion
collisions (see Refs.~\cite{Kharzeev:2015znc,Fukushima:2018grm} for
recent reviews).  Non-central collisions would produce a quark-gluon
plasma (or hot hadronic matter at lower collision energies) under
strong $\bB$ and large
$\bomega$~\cite{Tuchin:2013ie,Kharzeev:2015znc,Deng:2016gyh,STAR:2017ckg}.
Now, chiral transport phenomena in general have grown up to be one
major subject in not only heavy-ion collision but general
physics~\cite{Son:2012zy,Hattori:2019ahi}.  So far, $\bB$-induced and
$\bomega$-induced phenomena have been considered individually, but an
interplay between $\bB$ and $\bomega$ would be becoming more and more
attractive~\cite{Ebihara:2016fwa,Hattori:2016njk,Chen:2017xrj,Chernodub:2017mvp,Liu:2017zhl,Liu:2017spl,Wang:2017pje,Cao:2019ctl,Chen:2019tcp}.
Among various theoretical attempts, we shall pay our special attention
to Ref.~\cite{Hattori:2016njk} in which a finite density
$\propto\bomega\cdot\bB$ was discovered.  The authors in
Ref.~\cite{Hattori:2016njk} proposed redistribution of the vector
charge induced by $\bomega\cdot\bB$, which may carry a vital role in
magnetohydrodynamics where $\bB$ is too large to be regarded as the
first order in the gradient expansion~\cite{Son:2009tf}.  We will
reproduce this result as a benchmark test of our formulation and
discuss an intuitive picture to deepen our understanding on the
induced density.

In this work we shall intensively investigate fermion systems with
$\bB$ and analyze the mode decomposition.  We will express the
density, the chirality density, the vector current, and the
axial-vector current in terms of the angular momentum modes.  Our most
prominent finding is an elegant mode-by-mode equation between the
vector current $\bj$ and the chirality density $\rho_5$, as we already
mentioned above.  Notably the equation holds not only for the lowest
landau levels (LLLs) but also higher Landau levels for any mass, while
a similar equation between the axial-vector current $\bj_5$ and the
density $\rho$ is valid for the LLLs only.  The existence of such an
elegant relation reminds us of the fact that the CME is anomaly
protected.  We should also emphasize that the anomaly nature of the
CME appears from the LLL contribution, but the relation we found is
for all Landau levels.  We are proposing this mode decomposed CME as
an extended interpretation of the CME without $\mu_5$.  We note that
we can straightforwardly recover the familiar formula of the CME by
taking the mode sum with proper weights with $\mu_5$.

We can apply our mode decomposed formulation for rotating fermion
systems.  It is known that there is a reciprocal relationship between
$\bB$ and $\bomega$ but their microscopic descriptions are totally
different.  Once all angular modes are available, we can introduce
$\bomega$ in the weight factor in a form of the chemical potential
shift~\cite{Chen:2015hfc,Jiang:2016wvv}.  From this analogy between
the rotational and finite-density effects, we can say that states with
a certain angular momentum are reminiscent of those with a certain
particle number in the \textit{canonical ensemble}.  The Legendre
transformation gives the grand canonical ensemble with the chemical
potential, and a similar machinery would work to translate the angular
momentum modes into the \textit{grand canonical ensemble} with a
certain $\bomega$.  One might think that we just recapitulate known
calculations in the cylindrical coordinates, but our formulation can
reveal properties of rotating fermions especially under $\bB$ in an
intuitive manner.  As mentioned above, we will give a plain
explanation of induced density discussed in
Ref.~\cite{Hattori:2016njk}, through which subtleties about the
angular momentum associated with the magnetic field will be
manifested.

Finally, let us briefly mention a possibility that our calculation may
be directly applied to electron vortex beams in
optics~\cite{PhysRevLett.107.174802,PhysRevLett.112.134801,Bialynicki-Birula:2017moy,vanKruining:2017anw,PhysRevA.95.063812,PhysRevLett.121.043202,PhysRevLett.122.063201,Ivanov:2019vxe}.  Unlike the topological vortex in a superfluid, what is called the
fermion vortex is a fermion beam whose wave functions have a helical
phase structure.  For discussions from the view point of Berry phases, see
Refs.~\cite{Bliokh:2007ec,PhysRevLett.107.174802,Bliokh:2017sdz,ducharme2018fractional}.  It is also possible that light waves could have
orbital angular momenta as speculated first by Poynting and this has
been actually realized
experimentally~\cite{PhysRevA.45.8185,Bliokh:2014ara}.

The azimuthal dependence of electron vortex beams is
$e^{il\varphi}$ where $l\in\mathbb{Z}$ is a quantum number identified
with the orbital angular momentum.  Although a relativistic extension
of the electron vortex beam is a theoretically challenging
subject~\cite{Barnett:2017wrr}, nonrelativistic vortex beams have been
observed experimentally and their applications have become an exciting
research
field~\cite{Bliokh:2007ec,uchida2010generation,verbeeck2010production,mcmorran2011electron,clark2015controlling,harris2015structured,PhysRevLett.114.034801,Bliokh:2017uvr,doi:10.1063/1.4977879}.
In this work we express various physical quantities using the angular
momentum decomposed modes, and we can regard mode-by-mode expressions
as contributions from wave-functions of vortex beams.  Then, the mode
decomposed CME as we advocate in this work should be directly tested
with the vortex beams if the helicity of the vortex beams could be
projected out.  We can say, therefore, that we put forward an
experimental possibility to realize chirally-induced effects in
tabletop setups in the future.  It should be also emphasized that the
fermion vortex beams have a lot of future prospects.  The intrinsic
orbital angular momentum of the vortex beams could take an arbitrarily
large value in principle.  For example, it has been reported that
electron vortex beams up to $l\sim 1000$ (i.e., the orbital angular
momentum $\sim 1000\hbar$) have been experimentally
observed~\cite{doi:10.1063/1.4977879}.  Generally speaking, it is
difficult to design experiments with large vorticity $\bomega$ under
reasonable control, but such states with large $l$ would provide us
with an alternative and controllable probe to investigate rotating
fermions.

Although our theoretical background lies in QCD physics, we will
employ the convention of electrons and positrons in this work.  If
necessary, it would be easy to generalize our formulas for the purpose
of QCD studies in terms of quarks by adjusting electric charges.
Throughout this paper we employ the physical units of $\hbar=c=1$.

\section{Solutions of the Dirac Equation in the Cylindrical Coordinates}

We present explicit forms of the solutions of the Dirac equation in
the cylindrical coordinates with the quantum number $l$ corresponding
to the orbital angular momentum.  Those expressions also explain our
notations and conventions used throughout this paper.

The vacuum structures in quantum field theories should not be modified
solely by rotation once the boundary effects are properly taken into
account not to violate the
causality~\cite{Vilenkin:1980zv,Ebihara:2016fwa}.  In contrast, as
demonstrated in
Refs.~\cite{Liu:2017zhl,Liu:2017spl,Wang:2018zrn,Wang:2019nhd}, the
vacuum structures would change if the fermionic wave-functions are
localized by coupling to external environments and thus become
insensitive to the boundary effects.  In this work we shall utilize
the cylindrical coordinates, $(r,\varphi,z)$, under a constant
magnetic field, $\bB=B \be_z$, so that the Landau wave-functions are
exponentially localized on the transverse $(r,\varphi)$ plane.

Throughout this work we employ the Dirac representation for the
$\gamma$ matrices, i.e.,
\begin{equation}
  \gamma^0 = \begin{pmatrix} I & 0 \\ 0 & -I \end{pmatrix}\,,
  \qquad
  \gamma^i = \begin{pmatrix} 0 & \sigma_i \\ -\sigma_i & 0 \end{pmatrix}\,,
\end{equation}
where $I$ is the $2\times 2$ unit matrix and $\sigma_i$'s are the
Pauli matrices.  In this convention $\gamma_5$ takes a non-diagonal
form of
\begin{equation}
  \gamma_5 = \begin{pmatrix} 0 & I \\ I & 0 \end{pmatrix}\,.
\end{equation}
The solutions of the free Dirac equation with the electric charge
convention, $e<0$, read in the Dirac representation:
\begin{equation}
  \begin{split}
    & u^{(\uparrow)}_{n,l,k}(r,\varphi,z,t) \\
    & = \frac{e^{-i\varepsilon^{(\uparrow)}_{n,l,k} t + ikz}}
    {\sqrt{\varepsilon^{(\uparrow)}_{n,l,k}+m}}
    \begin{pmatrix}
      (\varepsilon^{(\uparrow)}_{n,l,k}+m)\Phi_{n,l}(\chi^2,\varphi) \\[0.5em]
      0 \\[0.2em]
      k\Phi_{n,l}(\chi^2,\varphi) \\[0.5em]
      i\sqrt{|e|B(2n\!+\!|l|\!+\!l\!+\!2)}\, \Phi_{n,l+1}(\chi^2,\varphi)
    \end{pmatrix}
  \end{split}
  \label{eq:up}
\end{equation}
for the state with the positive helicity, the positive energy, and (the
$z$ component of) the total angular momentum, $J_z = l+\frac{1}{2}$.
In the above expression $n$ and $l$ form the Landau level index and
$l$ and $s$ represent the $z$-component eigenvalues of the orbital and
the spin angular momenta, respectively.  The $z$-component of the
three momentum is denoted by $k$.  A mixture of $l$ and $l+1$ in the
spinor components stems from twofold realization of the same $J_z$ as
$l+\frac{1}{2}$ and $(l+1)-\frac{1}{2}$.  In the same way, the
negative helicity and positive energy states read:
\begin{equation}
  \begin{split}
    & u^{(\downarrow)}_{n,l,k}(r,\varphi,z,t) \\
    & = \frac{e^{-i\varepsilon^{(\downarrow)}_{n,l,k} t + ikz}}
    {\sqrt{\varepsilon^{(\downarrow)}_{n,l,k}+m}}
    \begin{pmatrix}
      0 \\[0.2em]
      (\varepsilon^{(\downarrow)}_{n,l,k}+m)\Phi_{n,l+1}(\chi^2,\varphi) \\[0.5em]
      -i\sqrt{|e|B(2n\!+\!|l+1|\!+\!l\!+\!1)}\, \Phi_{n,l}(\chi^2,\varphi)   \\[0.5em]
      -k\Phi_{n,l+1}(\chi^2,\varphi)
    \end{pmatrix}
  \end{split}
  \label{eq:down}
\end{equation}
for the same total angular momentum, $J_z=l+\frac{1}{2}$.  In these
expressions~\eqref{eq:up} and \eqref{eq:down} we introduced a
dimensionless variable $\chi^2 = \frac{1}{2}eB r^2$ and defined a
function $\Phi_{n,l}(\chi^2,\varphi)$ as
\begin{equation}
  \Phi_{n,l}(\chi^2,\varphi) = 
  \sqrt{\frac{n!}{(n+|l|)!}}\, e^{-\frac{1}{2}\chi^2}\,
  \chi^{|l|} L_n^{|l|}(\chi^2)\, e^{il\varphi} \,.
\end{equation}
It would be useful to make it clear how we chose the normalization of
the above function, that is., our convention is,
\begin{equation}
  \int \! dxdy\, |\Phi_{n,l}(\chi^2,\!\varphi)|^2
  = \frac{2\pi}{|eB|}\int \! d\chi^2 |\Phi_{n,l}(\chi^2,\!\varphi)|^2
  = \frac{2\pi}{|eB|}\,.
  \label{eq:Phi_norm}
\end{equation}
This Landau wave-function, representing a localized spatial profile,
has an exponential damping factor for large $r^2$.  For these states
of $u^{(\uparrow)}_{n,l,k}$ and $u^{(\downarrow)}_{n,l,k}$, the one
particle eigenenergies are given by
\begin{equation}
  \begin{cases}
  \varepsilon^{(\uparrow)}_{n,l,k} = \sqrt{|e|B(2n+|l|+l+2)+k^2+m^2}\,, \\[1em]
  \varepsilon^{(\downarrow)}_{n,l,k} = \sqrt{|e|B(2n + |l+1| + l+1)
    + k^2 + m^2} \,.
  \end{cases}
\label{eq:energy_p}
\end{equation}
We note that the normalization~\eqref{eq:Phi_norm} corresponds to the
following convention for the Dirac spinor normalization:
\begin{equation}
  \frac{|eB|}{2\pi} \int dxdy\,
  u^{(\uparrow,\downarrow)\dag}_{n,l,k} u^{(\uparrow,\downarrow)}_{n,l,k}
  = 2\varepsilon^{(\uparrow,\downarrow)}_{n,l,k} \,.
  \label{eq:spinornorm}
\end{equation}
This is reduced to the standard normalization in the limit of
$eB\to 0$, which can be understood from
$\int dxdy\,e^{-\chi^2}=2\pi/|eB|$.  In later discussions, in fact, we
will consider the $eB\to 0$ limit, which needs extra care about the
treatment of $2\pi/|eB|$ which is cut off by the transverse area
$S_\perp$ for $eB\to 0$.

The anti-particle solutions of the free Dirac equation in the magnetic
field with $J_z=l+\frac{1}{2}$ read:
\begin{equation}
  \begin{split}
    & v^{(\uparrow)}_{n,l,k}(r,\varphi,z,t) \\
    & = \frac{e^{i\bar{\varepsilon}^{(\uparrow)}_{n,l,k} t - ikz}}
    {\sqrt{\bar{\varepsilon}^{(\uparrow)}_{n,l,k}+m}}
    \begin{pmatrix}
      -i\sqrt{|e|B(2n\!+\!|l|\!-\!l)}\, \Phi_{n,-l-1}(\chi^2,\varphi) \\[0.5em]
      -k\Phi_{n,-l}(\chi^2,\varphi) \\[0.5em]
      0 \\[0.2em]
      (\bar{\varepsilon}^{(\uparrow)}_{n,l,k}+m)\Phi_{n,-l}(\chi^2,\varphi)
    \end{pmatrix}
  \end{split}
  \label{eq:anti-up}
\end{equation}
for the positive helicity anti-particle state.  The anti-particle
states with the negative helicity read:
\begin{equation}
  \begin{split}
    & v^{(\downarrow)}_{n,l,k}(r,\varphi,z,t) \\
    & = \frac{e^{i\bar{\varepsilon}^{(\downarrow)}_{n,l,k} t - ikz}}
    {\sqrt{\bar{\varepsilon}^{(\downarrow)}_{n,l,k}+m}}
    \begin{pmatrix}
      -k\Phi_{n,-l-1}(\chi^2,\varphi) \\[0.5em]
      -i\sqrt{|e|B(2n\!+\!|l+1|\!-\!l+1)}\, \Phi_{n,-l}(\chi^2,\varphi) \\[0.5em]
     - (\bar{\varepsilon}^{(\downarrow)}_{n,l,k}+m)\Phi_{n,-l-1}(\chi^2,\varphi) \\[0.5em]
      0
    \end{pmatrix}
  \end{split}
  \label{eq:anti-down}
\end{equation}
which carries the same total angular momentum, $J_z=l+\frac{1}{2}$.
The anti-particle eigenenergies are written as
\begin{equation}
  \begin{cases}
  \bar{\varepsilon}^{(\uparrow)}_{n,l,k} = \sqrt{|e|B(2n+|l|-l)+k^2+m^2}\,, \\[1em]
  \bar{\varepsilon}^{(\downarrow)}_{n,l,k} = \sqrt{|e|B(2n + |l+1| - l+1)
    + k^2 + m^2} \,.
  \end{cases}
\end{equation}
Now, for developing deeper understanding of magnetic properties, it is
crucially important to see where the lowest Landau levels (LLLs) lie
in this setup.  For particles with $e<0$ (or the electrons
particularly) we can immediately identify the LLLs as $(\downarrow)$
states with $n=0$ and $l\leq -1$ (i.e., $J_z < 0$), for which
$\varepsilon^{(\downarrow)}_{n=0,l\leq -1,k}=\sqrt{k^2+m^2}$.
For anti-particles (or the positrons), in the same way, the LLLs are
found to be $(\uparrow)$ states with $n=0$ and $l\geq 0$ (i.e.,
$J_z>0$), for which
$\bar{\varepsilon}^{(\uparrow)}_{n=0,l\geq 0,k}=\sqrt{k^2+m^2}$.
The range of $l$ is $(-\infty,\infty)$ but the LLLs span over
$-S_\perp |eB|/(2\pi) < J_z < 0$ for the $(\downarrow)$ particles and
$0<J_z < S_\perp|eB|/(2\pi)$ for the $(\uparrow)$ anti-particles for
sufficiently large $S_\perp |eB|/(2\pi)$ (see Ref.~\cite{Chen:2017xrj}
for details).

\begin{figure}
  \includegraphics[width=0.6\columnwidth]{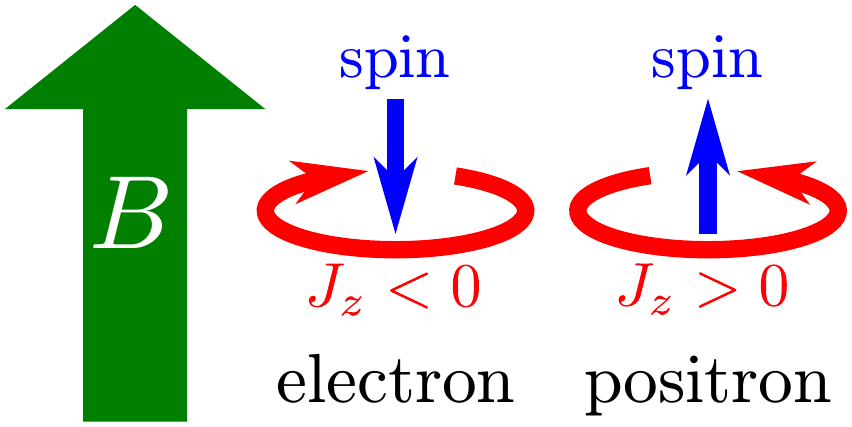}
  \caption{Schematic illustration of the LLLs for negatively charged
    particles (electrons) and positively charged anti-particles
    (positrons).  The preferred spin and orbital alignments are
    consistent with classical motion of charged particles.}
  \label{fig:LLL}
\end{figure}

We note that the preferred $(\uparrow,\downarrow)$ and the range of
$l$ for these LLLs are completely consistent with our intuition based
on classical physics;  the energy is lowered by the
spin alignment anti-parallel to $B$ for negatively charged particles,
while the spin alignment should be parallel to $B$ for positively charged
anti-particles.  The orbital dynamics is in accord to the Larmor
motion in classical electrodynamics and $J_z<0$ for electrons and
$J_z>0$ for positrons are naturally concluded, respectively.  For a
schematic illustration, see Fig.~\ref{fig:LLL}.

\section{Mode Decomposed Density Distributions}

We will consider mode-by-mode contributions to the expectation values
of the scalar (density), the pseudo-scalar (chirality density), the
vector, and the axial-vector operators.  We will later make use of the
results in this section to propose the mode decomposed version of the
CME{}.  Also we apply our results for a canonical formulation of
rotating fermions.

\subsection{Density and Chirality Density}

The fermion number density is given by
$\rho=\langle\hat{\psi}^\dag \hat{\psi}\rangle$, where $\hat{\psi}$
represents a Dirac field operator.  It is a standard procedure to
expand $\hat{\psi}$ in terms of the complete basis of
$u^{(\uparrow,\downarrow)}_{n,l,k}$ with particle annihilation
operators $\hat{a}^{(\uparrow,\downarrow)}_{n,l,k}$ and
$v^{(\uparrow,\downarrow)}_{n,l,k}$ with anti-particle
creation operators $\hat{b}^{(\uparrow,\downarrow)\dag}_{n,l,k}$ in
the second quantization method.
All annihilation and creation operators are normal-ordered.
Subtleties of the normal-ordered operators were discussed in
Refs.~\cite{Sheng:2017lfu,Dong:2020zci}.  In this work, however, we
consider one-particle states individually, so we have dropped
infinite classical numbers from the anti-commutation relation.
Then, we can express $\rho$ as a
linear superposition of different $(n,l,k)$ contributions, which we
can symbolically represent in the following form:
\begin{equation}
  \rho = \sumint_{n,l,k}\, ( \rho^{(\uparrow)}_{n,l,k} +\rho^{(\downarrow)}_{n,l,k}
  + \bar{\rho}^{(\uparrow)}_{n,l,k}
  + \bar{\rho}^{(\downarrow)}_{n,l,k} )\,,
\label{eq:netdensity}
\end{equation}
where the first two (and the last two) represent the particle (and the
anti-particle) contributions.  The sum-integral is a short-hand
representation of the phase space sum over $(n,l,k)$ with the proper
weight having the mass dimension three.  More explicitly, the phase
space integral is replaced as
\begin{equation}
  \int\frac{d^2 k_\perp}{(2\pi)^2} \;\Leftrightarrow\;
  \frac{1}{S_\perp} {\sum_{n,l}}' \;\to\;
  \frac{|eB|}{2\pi} \sum_{n,l} \,,
  \label{eq:replace}
\end{equation}
where $S_\perp$ is the transverse area and ${\sum}'$ denotes a
weighted sum that can reproduce the phase space integral in the zero
magnetic limit and the weight goes to $S_\perp |eB|/(2\pi)$ for
sufficiently large $S_\perp |eB|/(2\pi)$.  For a precise definition, see
Ref.~\cite{Chen:2017xrj};  we need to cope with a finite sized
boundary condition and this is beyond the current scope.

These expectation values of $\rho^{(\uparrow,\downarrow)}_{n,l,k}$ and
$\bar{\rho}^{(\uparrow,\downarrow)}_{n,l,k}$ depend on the state.  If
we take an expectation value with
$\hat{a}^{(\uparrow)\dag}_{n,l,k}|0\rangle$, we can immediately find the
density constituent, using the explicit solutions of the Dirac
equation, as
\begin{equation}
  \begin{split}
    \rho^{(\uparrow)}_{n,l,k}
    &= \frac{(\varepsilon^{(\uparrow)}_{n,l,k}+m)^2+k^2}
    {2\varepsilon^{(\uparrow)}_{n,l,k}(\varepsilon^{(\uparrow)}_{n,l,k}+m)}
  |\Phi_{n,l}|^2 \\
  &\qquad\qquad + \frac{|e|B(2n+|l|+l+2)}
  {2\varepsilon^{(\uparrow)}_{n,l,k}(\varepsilon^{(\uparrow)}_{n,l,k}+m)}|\Phi_{n,l+1}|^2 \,.
  \end{split}
\label{eq:rhoup}
\end{equation}
In the same way, we consider an expectation value corresponding to a
state, $\hat{a}^{(\downarrow)\dag}_{n,l,k}|0\rangle$, which turns out to be
\begin{equation}
  \begin{split}
    \rho^{(\downarrow)}_{n,l,k}
    &= \frac{(\varepsilon^{(\downarrow)}_{n,l,k}+m)^2+k^2}
    {2\varepsilon^{(\downarrow)}_{n,l,k}(\varepsilon^{(\downarrow)}_{n,l,k}+m)}
  |\Phi_{n,l+1}|^2 \\
  &\qquad\qquad + \frac{|e|B(2n+|l+1|+l+1)}
  {2\varepsilon^{(\downarrow)}_{n,l,k}(\varepsilon^{(\downarrow)}_{n,l,k}+m)}|\Phi_{n,l}|^2 \,.
  \end{split}
\label{eq:rhodown}
\end{equation}
Similarly, for the anti-particle contributions, states
$\hat{b}^{(\uparrow,\downarrow)\dag}_{n,l,k}|0\rangle$ lead to the
expectation values as given by
\begin{equation}
  \begin{split}
    \bar{\rho}^{(\uparrow)}_{n,l,k}
    &= -\frac{(\bar{\varepsilon}^{(\uparrow)}_{n,l,k}+m)^2+k^2}
    {2\bar{\varepsilon}^{(\uparrow)}_{n,l,k}(\bar{\varepsilon}^{(\uparrow)}_{n,l,k}+m)}
  |\Phi_{n,-l}|^2 \\
  &\qquad\qquad - \frac{|e|B(2n+|l|-l)}
  {2\bar{\varepsilon}^{(\uparrow)}_{n,l,k}(\bar{\varepsilon}^{(\uparrow)}_{n,l,k}+m)}|\Phi_{n,-l-1}|^2 \,,
  \end{split}
\label{eq:brhoup}
\end{equation}
and
\begin{equation}
  \begin{split}
    \bar{\rho}^{(\downarrow)}_{n,l,k}
    &= -\frac{(\bar{\varepsilon}^{(\downarrow)}_{n,l,k}+m)^2+k^2}
    {2\bar{\varepsilon}^{(\downarrow)}_{n,l,k}(\bar{\varepsilon}^{(\downarrow)}_{n,l,k}+m)}
  |\Phi_{n,-l-1}|^2 \\
  &\qquad\qquad - \frac{|e|B(2n+|l+1|-l+1)}
  {2\bar{\varepsilon}^{(\downarrow)}_{n,l,k}(\bar{\varepsilon}^{(\downarrow)}_{n,l,k}+m)}|\Phi_{n,-l}|^2 \,.
  \end{split}
\label{eq:brhodown}
\end{equation}
The overall minus sign of $\bar{\rho}$ comes from the anti-commutation relation of $\hat{b}$.
For our later quantitative discussions it is important to note that
the spatially integrated quantities (denoted by $\aver$ here) are,
corresponding to Eq.~\eqref{eq:Phi_norm}, quantized as
\begin{equation}
  \begin{split}
    \aver^{(\uparrow,\downarrow)}_{n,l,k}
    &= \int dxdy\, \rho^{(\uparrow,\downarrow)}_{n,l,k}
    = \frac{2\pi}{|eB|} \,,\\
    \bar{\aver}^{(\uparrow,\downarrow)}_{n,l,k}
    &= \int dxdy\, \bar{\rho}^{(\uparrow,\downarrow)}_{n,l,k}
    = -\frac{2\pi}{|eB|}
  \end{split}
  \label{eq:rhoint}
\end{equation}
for any $(n,l,k)$.

\begin{figure}
  \includegraphics[width=\columnwidth]{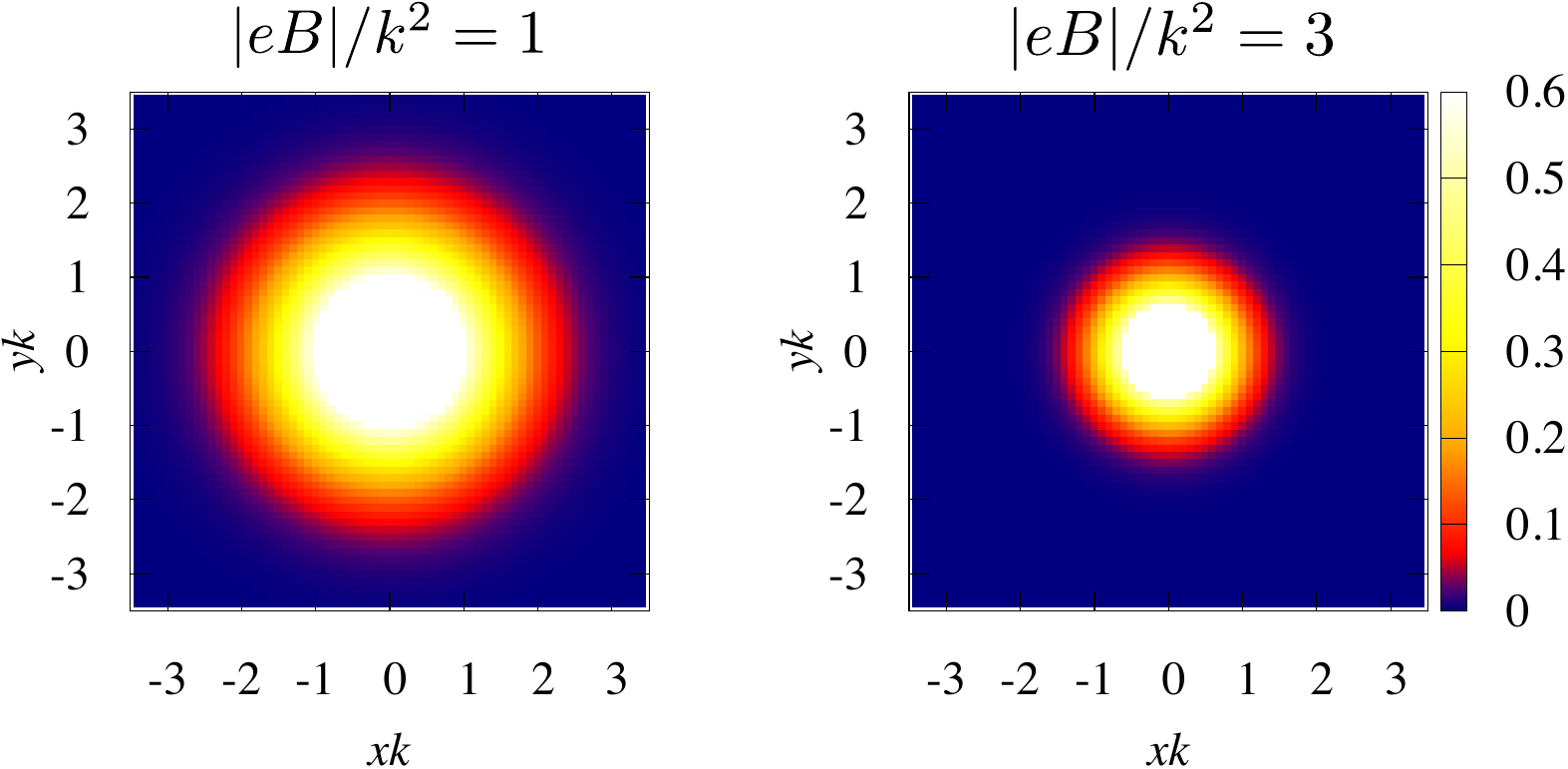}
  \caption{Density plot of $\rho^{(\downarrow)}_{n,l,k}$ from one LLL
    with $n=0$, $l=-1$ (i.e., $J_z=-\frac{1}{2}$) with all
    dimensionful quantities rescaled with $k$.  The mass is chosen as
    $m/k=1$.  The left panel is for the magnetic strength $|e|B/k^2=1$
    and the right one for $|e|B/k^2=3$.}
  \label{fig:ctBlm1}
\end{figure}

\begin{figure}
  \includegraphics[width=\columnwidth]{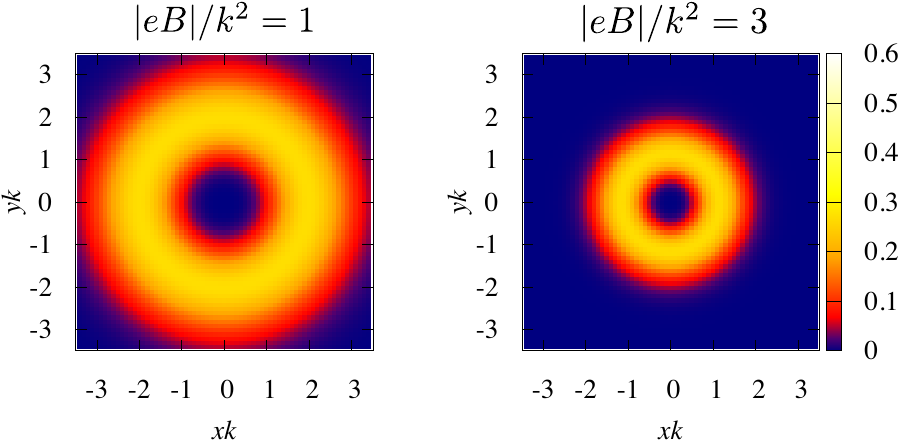}
  \caption{Density plot of $\rho^{(\downarrow)}_{n,l,k}$ from another LLL
    with $n=0$, $l=-3$ (i.e., $J_z=-\frac{5}{2}$) with all
    dimensionful quantities rescaled with $k$.  The mass is chosen as
    $m/k=1$.  The left panel is for the magnetic strength $|e|B/k^2=1$
    and the right one for $|e|B/k^2=3$.}
  \label{fig:ctBlm3}
\end{figure}

We make density plots for $\rho^{(\downarrow)}_{n,l,k}$ in
Figs.~\ref{fig:ctBlm1} and \ref{fig:ctBlm3} to visualize spatial
distributions of the $s$-wave and $d$-wave LLL states (there are
essentially the same plots shown in Ref.~\cite{PhysRevA.95.063812}).
In our assignment $l=-1$ for the $\downarrow$ spin corresponds to the
$s$-wave state with $J_z=-\frac{1}{2}$, so that the density is peaked
around the origin in Fig.~\ref{fig:ctBlm1}.  As the magnetic field
increases from the left panel to the right panel, the wave-function
becomes more localized near the origin, as is clear in both
Figs.~\ref{fig:ctBlm1} and \ref{fig:ctBlm3}.  Then, such a suppression
factor by $1/(eB)$ in Eq.~\eqref{eq:rhoint} is naturally understood
from reduction of distributed areas.  Figure~\ref{fig:ctBlm3} is a plot
for the $l=-3$ or $J_z=-\frac{5}{2}$ mode.  In this case the
centrifugal force makes the wave-function peaks farther from the origin
and there is a hollow around the origin.

Next, we shall consider the chirality density, i.e.,
$\rho_5=\langle\hat{\psi}^\dag \gamma_5 \hat{\psi}\rangle$, in the
same way, which can be symbolically decomposed again as
\begin{equation}
  \rho_5 = \sumint_{n,l,k}\, ( \rho^{(\uparrow)}_{5\;n,l,k} +\rho^{(\downarrow)}_{5\;n,l,k}
  + \bar{\rho}^{(\uparrow)}_{5\;n,l,k} + \bar{\rho}^{(\downarrow)}_{5\;n,l,k} )\,.
\end{equation}
Here, simple calculations immediately lead to the following expressions:
\begin{equation}
  \rho^{(\uparrow)}_{5\;n,l,k} = \frac{k}{\varepsilon^{(\uparrow)}_{n,l,k}}
  |\Phi_{n,l}|^2 \,, \quad
  \rho^{(\downarrow)}_{5\;n,l,k} =
  -\frac{k}{\varepsilon^{(\downarrow)}_{n,l,k}}
  |\Phi_{n,l+1}|^2
\end{equation}
for $\hat{a}^{(\uparrow,\downarrow)\dag}_{n,l,k}|0\rangle$ states and
\begin{equation}
  \bar{\rho}^{(\uparrow)}_{5\;n,l,k} = \frac{k}{\bar{\varepsilon}^{(\uparrow)}_{n,l,k}}
  |\Phi_{n,-l}|^2 \,, \quad
  \bar{\rho}^{(\downarrow)}_{5\;n,l,k} =
  -\frac{k}{\bar{\varepsilon}^{(\downarrow)}_{n,l,k}}
  |\Phi_{n,-l-1}|^2
\end{equation}
for $\hat{b}^{(\uparrow,\downarrow)\dag}_{n,l,k}|0\rangle$ states.
These are surprisingly simple expressions as compared to counterparts
of the fermion number density.  Unlike the fermion density, we see
that the net chirality is vanishing after taking the mode sum over
$k$.  This is because the combination of $(\uparrow,\downarrow)$ and
$k$ uniquely fixes whether the chirality is positive or negative.
Usually for massless fermions the chirality is determined by the spin
and the momentum directions;  in the present setup $k$ is nothing but
the momentum direction and $(\uparrow,\downarrow)$ corresponds to the
spin direction.  Supposing that states with a particular $k\neq 0$ are
prepared, LLLs are states with the largest chirality, i.e.,
\begin{equation}
  \begin{split}
    \aver^{(\downarrow)}_{5,{\rm LLL}}
    &=\int dxdy\, \rho_{5,{\rm LLL}}^{(\downarrow)}
      = -\frac{2\pi}{|eB|}  \frac{k}{\varepsilon_k} \,,\\
    \bar{\aver}^{(\uparrow)}_{5,{\rm LLL}}
    &= \int dxdy\, \bar{\rho}_{5,{\rm LLL}}^{(\uparrow)}
      = \frac{2\pi}{|eB|}  \frac{k}{\varepsilon_k}
  \end{split}
\end{equation}
after the spatial integration, where $\varepsilon_k=\sqrt{k^2+m^2}$.
In the massless limit of $m\to 0$, we see that $k/\varepsilon_k$
reduces to the sign function of $k$.

In this paper we would not plot
$\rho_{5\; n,l,k}^{(\uparrow,\downarrow)}$ nor
$\bar{\rho}_{5\; n,l,k}^{(\uparrow,\downarrow)}$, for they look
indistinguishably similar to Figs.~\ref{fig:ctBlm1} and
\ref{fig:ctBlm3} on the qualitative level.

\subsection{Vector and Axial Vector Currents}

We can further proceed to the vector and the axial vector currents,
that is,
$j^z = \langle \hat{\psi}^\dag \gamma^0\gamma^z \hat{\psi}\rangle$
and
$j_5^z = \langle \hat{\psi}^\dag \gamma^0\gamma^z \gamma_5 \hat{\psi}\rangle$
mode by mode.  Interestingly, for the
vector components, we find such simple expressions as
\begin{equation}
  j^{z(\uparrow)}_{n,l,k} = \frac{k}{\varepsilon^{(\uparrow)}_{n,l,k}}
  |\Phi_{n,l}|^2 \,,\quad
  j^{z(\downarrow)}_{n,l,k} = \frac{k}{\varepsilon^{(\downarrow)}_{n,l,k}}
  |\Phi_{n,l+1}|^2
\end{equation}
for $\hat{a}^{(\uparrow,\downarrow)\dag}_{n,l,k}|0\rangle$ states and
\begin{equation}
  \bar{j}^{z(\uparrow)}_{n,l,k} = -\frac{k}{\bar{\varepsilon}^{(\uparrow)}_{n,l,k}}
  |\Phi_{n,-l}|^2 \,,\quad
  \bar{j}^{z(\downarrow)}_{n,l,k} = -\frac{k}{\bar{\varepsilon}^{(\downarrow)}_{n,l,k}}
  |\Phi_{n,-l-1}|^2
\end{equation}
for $\hat{b}^{(\uparrow,\downarrow)\dag}_{n,l,k}|0\rangle$ states.
Here, it is a quite interesting and profound observation that the
following relations should hold:
\begin{equation}
  \begin{split}
  j^{z(\uparrow)}_{n,l,k} = \rho^{(\uparrow)}_{5\,n,l,k}\,,\qquad
  j^{z(\downarrow)}_{n,l,k} = -\rho^{(\downarrow)}_{5\,n,l,k} \,,\\
  \bar{j}^{z(\uparrow)}_{n,l,k} = -\bar{\rho}^{(\uparrow)}_{5\,n,l,k}\,,\qquad
  \bar{j}^{z(\downarrow)}_{n,l,k} = \bar{\rho}^{(\downarrow)}_{5\,n,l,k}
  \end{split}
  \label{eq:jrhorel}
\end{equation}
for \textit{any}
$\hat{a}^{(\uparrow,\downarrow)\dag}_{n,l,k}|0\rangle$ and
$\hat{b}^{(\uparrow,\downarrow)\dag}_{n,l,k}|0\rangle$ states.  We
would emphasize that the above simple proportionality holds even
beyond the LLLs.

The axial vector part is a little more complicated.  After several line
calculations we arrive at the following expressions;
\begin{equation}
  \begin{split}
    j^{z(\uparrow)}_{5\,n,l,k}
    &= \frac{(\varepsilon^{(\uparrow)}_{n,l,k}+m)^2+k^2}
    {2\varepsilon^{(\uparrow)}_{n,l,k}(\varepsilon^{(\uparrow)}_{n,l,k}+m)}
  |\Phi_{n,l}|^2 \\
  &\qquad\qquad - \frac{|e|B(2n+|l|+l+2)}
  {2\varepsilon^{(\uparrow)}_{n,l,k}(\varepsilon^{(\uparrow)}_{n,l,k}+m)}|\Phi_{n,l+1}|^2
  \end{split}
\end{equation}
for $\hat{a}^{(\uparrow)\dag}_{n,l,k}|0\rangle$ states and
\begin{equation}
  \begin{split}
    j^{z(\downarrow)}_{5\,n,l,k}
    &= -\frac{(\varepsilon^{(\downarrow)}_{n,l,k}+m)^2+k^2}
    {2\varepsilon^{(\downarrow)}_{n,l,k}(\varepsilon^{(\downarrow)}_{n,l,k}+m)}
  |\Phi_{n,l+1}|^2 \\
  &\qquad\qquad + \frac{|e|B(2n+|l+1|+l+1)}
  {2\varepsilon^{(\downarrow)}_{n,l,k}(\varepsilon^{(\downarrow)}_{n,l,k}+m)}|\Phi_{n,l}|^2
  \end{split}
\end{equation}
for $\hat{a}^{(\downarrow)\dag}_{n,l,k}|0\rangle$ states.  In the same way,
\begin{equation}
  \begin{split}
    \bar{j}^{z(\uparrow)}_{5\,n,l,k}
    &= \frac{(\bar{\varepsilon}^{(\uparrow)}_{n,l,k}+m)^2+k^2}
    {2\bar{\varepsilon}^{(\uparrow)}_{n,l,k}(\bar{\varepsilon}^{(\uparrow)}_{n,l,k}+m)}
  |\Phi_{n,-l}|^2 \\
  &\qquad\qquad - \frac{|e|B(2n+|l|-l)}
  {2\bar{\varepsilon}^{(\uparrow)}_{n,l,k}(\bar{\varepsilon}^{(\uparrow)}_{n,l,k}+m)}|\Phi_{n,-l-1}|^2
  \end{split}
\end{equation}
for $\hat{b}^{(\uparrow)\dag}_{n,l,k}|0\rangle$ states and
\begin{equation}
  \begin{split}
    \bar{j}^{z(\downarrow)}_{5\,n,l,k}
    &= -\frac{(\bar{\varepsilon}^{(\downarrow)}_{n,l,k}+m)^2+k^2}
    {2\bar{\varepsilon}^{(\downarrow)}_{n,l,k}(\bar{\varepsilon}^{(\downarrow)}_{n,l,k}+m)}
  |\Phi_{n,-l-1}|^2 \\
  &\qquad\qquad + \frac{|e|B(2n+|l+1|-l+1)}
  {2\bar{\varepsilon}^{(\downarrow)}_{n,l,k}(\bar{\varepsilon}^{(\downarrow)}_{n,l,k}+m)}|\Phi_{n,-l}|^2
  \end{split}
\end{equation}
for $\hat{b}^{(\downarrow)\dag}_{n,l,k}|0\rangle$ states.  One might
think that the above expressions look similar to previous
Eqs.~\eqref{eq:rhoup} and \eqref{eq:rhodown}, but the relative sign
between two terms is different.  Therefore, an elegant mode-by-mode
relation like Eq.~\eqref{eq:jrhorel} cannot generally exist for the
axial vector current.

Once we perform the spatial integration, we can sort out two terms
into one, which yields:
\begin{equation}
  \begin{split}
  \avej^{z(\uparrow)}_{5\,n,l,k} &= \int dxdy\, j^{z(\uparrow)}_{5\,n,l,k} \\
  &= \biggl[ 1-\frac{|e|B(2n+|l|+l+2)}{\varepsilon^{(\uparrow)}_{n,l,k}
    (\varepsilon^{(\uparrow)}_{n,l,k} + m)} \biggr] \;
  \aver^{(\uparrow)}_{n,l,k} \,,
  \end{split}
  \label{eq:j5upinteg}
\end{equation}
and
\begin{equation}
  \begin{split}
  \avej^{z(\downarrow)}_{5\,n,l,k} &= \int dxdy\, j^{z(\downarrow)}_{5\,n,l,k} \\
  &= -\biggl[ 1-\frac{|e|B(2n+|l+1|+l+1)}{\varepsilon^{(\downarrow)}_{n,l,k}
    (\varepsilon^{(\downarrow)}_{n,l,k} + m)} \biggr] \;
  \aver^{(\downarrow)}_{n,l,k} \,.
  \end{split}
  \label{eq:j5downinteg}
\end{equation}
Any further simplification is, however, impossible unless we limit
ourselves to the LLLs.  Indeed, for the LLLs only, a simple relation
realizes as follows:
\begin{equation}
  j^{z(\downarrow)}_{5\,{\rm LLL}} = -\rho^{(\downarrow)}_{\rm LLL}\,,\qquad
  \bar{j}^{z(\uparrow)}_{5\,{\rm LLL}} = -\bar{\rho}^{(\uparrow)}_{\rm LLL}
\label{eq:ajrhorel}
\end{equation}
for \textit{any} $m$, for the second terms drop off.  We note that
only the $(\downarrow)$ states have the LLLs for negatively charged
particles and there is no counterpart relation for the $(\uparrow)$
particle states.  The same can be said about the anti-particle LLL
states.

\section{Application to the Chiral Magnetic and Related Effects}

The CME is characterized by a finite vector current induced by a
nonzero chirality under external magnetic fields.  The most compact
representation of the CME relies on the chiral chemical potential,
$\mu_5$ and the formula is $\bj=\mu_5/(2\pi^2) e\bB$ (where our $\bj$
is not an electric current but a vector current).  Theoretically
speaking, it would be desirable to define the CME without resorting to
$\mu_5$ that is a troublesome object.  A common alternative is a
parity-odd background $\sim \bE\cdot\bB$ and, here, we are proposing a
novel picture of the CME by means of the mode decomposition.

\subsection{Recovery of the Chiral Magnetic Effect}

We first discuss how to derive the well-known formula of the ordinary
CME with $\mu_5$ within the present framework.  In the presence of
$\mu_5$ the wave-functions we presented before are no longer
eigenstates and the eigenenergies should be reconsidered.
Generally speaking, $\mu_5$ dependence is involved (but known; see
Ref.~\cite{Fukushima:2008xe}), and yet, the $(\uparrow,\downarrow)$
parts give chirality contributions with equal weights due to
degeneracy with incremented $n$ and conversion by $k\to -k$.  Then,
the current contributions cancel out due to an extra minus sign in
Eq.~\eqref{eq:jrhorel}, so that only the LLL contribution survives, as
is the case in the standard CME computation.  Now, we further simplify
the calculation by taking the $m\to0$ limit.  We see that the LLL states
have definite chirality for $m=0$ and we do not have to re-diagonalize
the Hamiltonian.  Taking the spatial average, $(1/S_\perp)\int dxdy$,
we can then express the total current as
\begin{align}
  & j^z = j^{z(\downarrow)}_{\rm LLL} + \bar{j}^{z(\uparrow)}_{\rm LLL}
        = -\rho^{(\downarrow)}_{5,{\rm LLL}}
        - \bar{\rho}^{(\uparrow)}_{5,{\rm LLL}} \notag\\
      &= -\biggl( \frac{1}{S_\perp} \frac{2\pi}{|eB|} \biggr)
         \frac{|eB|}{2\pi} 
        \biggl( \frac{S_\perp |eB|}{2\pi}\biggr)
        \int\frac{dk}{2\pi}  \notag\\
      & \times \biggl[ -\text{sgn}(k) \,
        \theta(\mu_5\text{sgn}(k) \!-\! |k|)
        + \text{sgn}(k) \, \theta(-\mu_5\text{sgn}(k) \!-\! |k|) \biggr] \notag\\
      &= \frac{|eB| \mu_5}{2\pi^2} \,,
  \label{eq:cme}
\end{align}
where the factor in the first parentheses comes from the spatial
average on the wave-functions, the second factor by $|eB|/(2\pi)$ is
from the phase space weight in Eq.~\eqref{eq:replace}, and the third
factor is from the $J_z$ sum (degeneracy).  We again note that we
explicitly took the $m=0$ limit just for computational simplicity, but
the CME itself is insensitive to $m$.

\subsection{Mode Decomposed Chiral Magnetic Effect}

Next, we shall discuss the mode decomposed realization of the CME,
which is one of the central results in this work.  Before that, here,
let us point out that Eq.~\eqref{eq:jrhorel} explains why we could
find \textit{no} simple relation between $j^z$ and $\rho_5$ until we
make the helicity decomposition.  That is, according to
Eq.~\eqref{eq:jrhorel},
$j_{n,l,k}^z = j_{n,l,k}^{z(\uparrow)} + j_{n,l,k}^{z(\downarrow)}
= \rho_{5\;n,l,k}^{(\uparrow)} - \rho_{5\;n,l,k}^{(\downarrow)}$,
while the chirality is
$\rho_{5\;n,l,k} = \rho_{5\;n,l,k}^{(\uparrow)} + \rho_{5\;n,l,k}^{(\downarrow)}$,
and they cannot be equated.  Thus, the $(\uparrow,\downarrow)$
projection is essential to establish such a simple relation.

Historically, the CME was first addressed in terms of $\rho_5$ in
Ref.~\cite{Kharzeev:2007jp} (denoted differently in the original
literature) in the context of the heavy-ion collisions, and then
$\rho_5$ gave way to $\mu_5$ since Ref.~\cite{Fukushima:2008xe}, and
now we are turning back to the original description with $\rho_5$.
The point is that data from the heavy-ion collisions are convoluted
with all time evolution and spatial fluctuations and the helicity
decomposed measurement is unfeasible.  With electron vortex beams for
example, however, it may be possible to prepare wave-functions with
either $(\uparrow)$ or $(\downarrow)$ projected out.  If we do so, we
do no longer have to introduce $\mu_5$ but we can simply refer to
Eq.~\eqref{eq:jrhorel} as the CME;  more specifically to distinguish
from the conventional CME, we may well call our Eq.~\eqref{eq:jrhorel}
the \textit{mode decomposed chiral magnetic effect}.

Let us reiterate the key expressions:
\begin{equation}
  j^{z(\uparrow)}_{n,l,k} = \rho^{(\uparrow)}_{5\;n,l,k} \,,\qquad
  j^{z(\downarrow)}_{n,l,k} = -\rho^{(\downarrow)}_{5\;n,l,k}
  \label{eq:mdcme}
\end{equation}
for the particle states.  It is worth emphasizing that this relation
is not only for the LLL states but for any $(n,l,k)$ and any $m$.
This feature makes a sharp contrast to the ordinary CME in which only
the LLL contribution survives, as we already discussed.  We would
stress that this is a great advantage experimentally since
Eq.~\eqref{eq:mdcme} holds for arbitrary (even vanishing!) magnetic
field and fermion mass.  In other words, in the ordinary CME, the role
of the magnetic field is to separate $(\uparrow,\downarrow)$ states.
Therefore, the magnetic field would be anyway needed for such a
purpose of the helicity decomposition in real setups.

One might think that Eq.~\eqref{eq:mdcme} invokes an analogy to the
(1+1)-dimensional system, i.e., the Dirac matrices algebraically
satisfy:
\begin{equation}
  \gamma^\mu \gamma_5 = -\varepsilon^{\mu\nu}\gamma_\nu\,,
\end{equation}
in (1+1)-dimensional theories, where our convention is
$\varepsilon^{01}=-\varepsilon^{10}=1$.  Then, the above identity is
immediately translated into
\begin{equation}
  j^z_{\rm LLL} = -\rho_{5\;\rm LLL} \,, \qquad
  j^z_{5\;\rm LLL} = -\rho_{\rm LLL}
  \label{eq:onedim}
\end{equation}
for the effectively (1+1)-dimensional LLLs in the case with negatively
charged particles.  These relations, however, make sense only when the
magnetic field is strong enough to justify the LLL approximation.  In
this sense Eq.~\eqref{eq:mdcme} keeps more physical contents including
higher LL correspondence.

\subsection{Contrast to the Chiral Separation Effect}

Now it would be instructive to consider the CSE here.  Unlike the CME,
the CSE is mass dependent, which was recognized since one of the
earliest works~\cite{Metlitski:2005pr}.  In the CSE an axial vector
current arises in finite-density matter in response to an external
magnetic field.  The CSE formula takes a simple form only in the
massless limit, and in fact, our Eqs.~\eqref{eq:j5upinteg} and
\eqref{eq:j5downinteg} can get simplified \textit{for the $m=0$ case only} as
\begin{equation}
  \avej^{z(\uparrow)}_{5\,n,l,k}
  = \frac{k^2}{\varepsilon^{(\uparrow)\,2}_{n,l,k}}
  \aver^{(\uparrow)}_{n,l,k} \,, \qquad
  \avej^{z(\downarrow)}_{5\,n,l,k}
  = -\frac{k^2}{\varepsilon^{(\downarrow)\,2}_{n,l,k}}
  \aver^{(\downarrow)}_{n,l,k} \,.
\label{eq:j5massless}
\end{equation}
In the same way as the CME derivation, with incremented $n$, the
energy dispersion relations for $(\uparrow,\downarrow)$ become
degenerated at finite $\mu$, and only the current contributions from
the LLLs survive.  From almost the same procedure as
Eq.~\eqref{eq:cme}, using Eq.~\eqref{eq:ajrhorel}, we can recover the
CSE as follows:
\begin{align}
  j_5^z &= j^{z(\downarrow)}_{5,{\rm LLL}} +
          \bar{j}^{z(\uparrow)}_{5,{\rm LLL}}
          = -\rho^{(\downarrow)}_{\rm LLL} -
          \bar{\rho}^{(\uparrow)}_{\rm LLL} \notag\\
          &= -\frac{|eB|}{2\pi}
  \int\frac{dk}{2\pi} \,
        \theta(\mu - |k|) = -\frac{|eB| \mu}{2\pi^2} \,.
  \label{eq:cse}
\end{align}
In this case, unlike Eq.~\eqref{eq:cme}, there is no contribution from
the anti-particle at zero temperature.  The overall minus sign is
intuitively understandable;  the axial vector current is identifiable
with the spin, and the particle LLL states have the spin anti-parallel
to the magnetic field as sketched in Fig.~\ref{fig:LLL}.

\section{Application to Rotating Fermions}

The mode decomposed formulation turns out to be insightful for us to
acquire intuitive understanding for rotational systems as well.  We shall
demonstrate concrete calculations and develop a perspective different
from the ordinary method.  To this end, first, we make a quick
overview of the conventional field-theoretical treatment of rotating
fermions.  In the rotating frame the Hamiltonian is modified into a
cranking form, which is analogous to finite density effects as pointed
out in Ref.~\cite{Chen:2015hfc}.  Bearing this analogy in mind, we
will introduce a working concept of ``grand canonical'' and
``canonical'' formulation of rotation.

\subsection{Rotating Fermions}

In the rotating frame we should incorporate a nontrivial metric as
well as vierbeins for spinors.  The covariant derivative in the Dirac
equation then involves the spin connection.  After some calculations
we can find that the free Dirac equation is given simply by
\begin{equation}
  \Bigl[ \gamma^0 ( iD_0 - \bomega \cdot \hat{\bJ} )
  -\bgamma\cdot i \bD - m \bigr] \psi = 0 \,,
  \label{eq:mod_Dirac}
\end{equation}
where the covariant derivative is $D_\mu=\partial_\mu + ieA_\mu$ with
our convention $e<0$ (for electrons), and $(\gamma^0, \bgamma)$ are defined in Dirac representation.  Here, $\hat{\bJ}$ is the total
angular momentum operator defined as
$\hat{\bJ}=\hat{\bl}+\hat{\bs}=\bx\times i\bnabla + \bSigma$ using the
spin matrix, $\bSigma=\frac{i}{4}\bgamma\times\bgamma$.  In the above
$\bomega$ denotes the angular velocity vector, which is taken along
the $z$ direction, i.e., $\bomega=\omega\be_z$, without loss of
generality.

\subsection{``Grand Canonical'' vs ``Canonical'' Formulations
of Rotation}

The modification in Eq.~\eqref{eq:mod_Dirac} clearly indicates that
the one-particle energy, $\varepsilon^{(\uparrow,\downarrow)}_{n,l,k}$
is shifted by a rotation-induced ``chemical potential'' as
\begin{equation}
  \varepsilon^{(\uparrow,\downarrow)}_{n,l,k} \;\to\;
  \varepsilon^{(\uparrow,\downarrow)}_{n,l,k}
  - \mu_{{\rm eff}\;l}\,,\qquad
  \mu_{{\rm eff}\;l} = \bomega\cdot \bJ = \omega \Bigl(l+\frac{1}{2}\Bigr)\,.
  \label{eq:shift_one_j}
\end{equation}
It is quite natural that a finite $\omega>0$ energetically favors
states with larger $J_z>0$.  In many-body systems the total energy
$E_{\rm total}$ would be thus shifted as
\begin{equation}
  E_{\rm total} \;\to\; E_{\rm total} - \omega J_{\rm total} \,,
  \label{eq:shift_J}
\end{equation}
where $J_{\rm total}$ represents the total angular momentum.  Such an
energy shift is reminiscent of the situation at finite chemical
potential, that is,
\begin{equation}
  E_{\rm total} \;\to\; E_{\rm total} - \mu N_{\rm total}\,.
  \label{eq:shift_N}
\end{equation}
Here, $N_{\rm total}$ is the total particle number.  We see prominent
similarity between Eqs.~\eqref{eq:shift_J} and \eqref{eq:shift_N}.
This finite density analogy of the rotational effect has been already
pointed out in Ref.~\cite{Chen:2015hfc}, which has inspired a phase
diagram with the finite density axis replaced with the angular
velocity~\cite{Jiang:2016wvv}.

For finite density systems we know that $\mu>0$ in the grand canonical
ensemble plays a role of pressure to induce a finite expectation value
of density, i.e., $\langle N_{\rm total}\rangle>0$.  Alternatively, we
can adopt the canonical ensemble picture and realize a finite density
system by forcing $N_{\rm total}>0$ directly.  In the thermodynamic
limit (i.e., the limit of infinite volume and particle number) the
grand canonical and the canonical descriptions should be completely
equivalent.

For the rotating fermionic system the above correspondence of
$\mu\to\omega$ and $N_{\rm total}\to J_{\rm total}$ implies a
``canonical'' alternative with which $J_{\rm total}$ is directly
imposed instead of using $\omega\neq 0$.  In fact, in finite size
systems like nuclei treated in nuclear structure theory, it is a
common and well established method to deal with rotation of deformed
nuclei by performing the angular momentum projection.  Indeed the
angular momentum projection is required to restore rotational symmetry
spontaneously broken by deformation.  This analogy to the canonical
ensemble is, however, not perfect and there are several crucial
differences.  The most important difference comes from the fact that
we cannot take the thermodynamic limit for rotating systems;
otherwise, the causality bound would be violated.  The next important
difference is that the one particle energy shift in
Eq.~\eqref{eq:shift_one_j} is not a constant but changes with $J_z$,
while the density shift is just a constant by
$\varepsilon\to\varepsilon-\mu$.  Since the fermion occupation should
be one per phase space, fermions occupy low energy states up to
$\varepsilon < \mu$.  In contrast, in the rotational case such
occupied states with
$\varepsilon^{(\uparrow,\downarrow)}_{n,l,k} < \omega J_z$ are not
necessarily limited to low energy states but high angular momentum
states may be possibly accumulated near the boundary edges of the
finite size system~\cite{Ebihara:2016fwa}.  Therefore, unlike the
finite density case, the canonical approach to rotating fermionic
systems may exhibit qualitative differences from what is expected from
the conventional ``grand canonical'' description at $\omega\neq 0$.
Still, a superposition of all ``canonical'' states with appropriate
($\omega$ dependent) weight factors would recover the conventional
results obtained in the rotating frame at $\omega\neq 0$.

\subsection{Rotation Induced Density}

As an example to show how to utilize our expressions for rotating
fermions, let us consider the following problem.  Without rotation and
chemical potential the particle and the anti-particle contributions
cancel out in Eq.~\eqref{eq:netdensity}.  The interesting question is;
what would happen for $\omega\neq 0$ and $eB\neq 0$ ?.  Then, for the
particles and the anti-particles, an effective chemical potential is
induced and the distribution functions should be
$\theta(\omega J_z - \varepsilon^{(\uparrow,\downarrow)}_{n,l,k})$ and
$\theta(\omega J_z - \bar{\varepsilon}^{(\uparrow,\downarrow)}_{n,l,k})$,
respectively, at zero temperature.

For the calculation of the density $\rho$, as discussed in
Ref.~\cite{Hattori:2016njk}, only the LLL contribution remains finite
up to the linear order in $\omega$.  For the LLLs, then, $J_z<0$ for
$\omega > 0$ is chosen out for the negatively charged particles, and
$J_z>0$ is for the anti-particles.  It is clear from the above
step-function distribution functions that only the anti-particle LLL
states make nonzero contributions.  Therefore, the density is
attributed solely to the anti-particle LLL states with $\omega J_z
>0$, and this is along the direction of the anti-particle LLL motion
as illustrated in Fig.~\ref{fig:LLL}.  To summarize, for $m=0$, we can
express the spatially averaged density as follows:
\begin{align}
  \rho &= -\biggl( \frac{1}{S_\perp} \frac{2\pi}{|eB|} \biggr)
  \frac{|eB|}{2\pi} \sum_{J_z>0}^{S_\perp |eB|/(2\pi)}
  \int\frac{dk}{2\pi}\,
         \theta(\omega J_z - |k|) \notag\\
       &= -\frac{\omega}{\pi S_\perp} \sum_{J_z}^{S_\perp |eB|/(2\pi)}
         \Bigl( l+\frac{1}{2} \Bigr) \notag\\
       &= -\frac{\omega |eB|}{4\pi^2}
         + \text{(orbital part)} \,.
\end{align}
Therefore, the spin contribution to an induced density reads:
\begin{equation}
  \rho_{\rm spin} = -\frac{\omega |eB|}{4\pi^2} \,,
  \label{eq:inducedrho}
\end{equation}
which correctly coincides with Ref.~\cite{Hattori:2016njk}.  The
orbital part may look proportional to $S_\perp$ which diverges in the
thermodynamic limit.  This pathological behavior arises from the
magnetic contribution to the angular momentum and we need careful
treatments of the \textit{canonical} and the \textit{kinetic} orbital
angular momenta~\footnote{This important observation was pointed out
  to us by Kazuya~Mameda.}.  They are different by
$|eB|r^2/2$~\cite{PhysRevLett.113.240404} and this extra contribution
(coming from the Poynting vector) cancels the orbital part.  We will
present more detailed and explicit discussions on this subtle but
interesting point in another publication.

\subsection{Chiral Vortical Effect}

The CVE is characterized by an axial vector current induced in matter
at $\omega\neq 0$.  The current can exist at finite value of either
temperature or density, and we will consider the finite density
situation only in this paper.  Then, the CVE formula should be;
$\bj_5 = \mu^2/(2\pi^2) \bomega$ in the case of $m=0$ and $|eB|=0$.
The coexistence of finite magnetic field may change the formula and it
would be an intriguing question to generalize the CVE formula to the
finite magnetic case.  The extra magnetic contribution to the orbital
angular momentum is, however, a subtle problem and we will discuss it
in another publication.

In the massless limit the axial vector current (after the spatial
average) is given by
\begin{align}
  j_5^z &= \biggl( \frac{1}{S_\perp}\frac{2\pi}{|eB|} \biggr)
          \frac{|eB|}{2\pi} \sum_{n,l} \int \frac{dk}{2\pi} \biggl[
  \frac{k^2}{\varepsilon^{(\uparrow)2}_{n,l,k}}
  \theta(\omega J_z + \mu - \varepsilon^{(\uparrow)}_{n,l,k}) \notag\\
        &\quad - \frac{k^2}{\varepsilon^{(\downarrow)2}_{n,l,k}}
          \theta(\omega J_z + \mu -
          \varepsilon^{(\downarrow)}_{n,l,k})
          + \frac{k^2}{\bar{\varepsilon}^{(\uparrow)2}_{n,l,k}}
          \theta(\omega J_z - \mu -
          \bar{\varepsilon}^{(\uparrow)}_{n,l,k}) \notag\\
         &\quad - \frac{k^2}{\bar{\varepsilon}^{(\downarrow)2}_{n,l,k}}
          \theta(\omega J_z - \mu -
          \bar{\varepsilon}^{(\downarrow)}_{n,l,k}) \biggr]\,.
\end{align}
One might think that only the LLLs remain with incremented $n$, but in
this case $J_z$ enters the step functions and the cancellation is
incomplete.  Namely, the LLLs give:
\begin{align}
  j_{5,{\rm LLL}}^z &= \frac{1}{S_\perp}
  \biggl[ -\sum_{J_z<0}^{-S_\perp |eB|/(2\pi)} \int \frac{dk}{2\pi}
                                 \theta(\omega J_z + \mu - |k|) \notag\\
  &\qquad\qquad + \sum_{J_z>0}^{S_\perp |eB|/(2\pi)} \int \frac{dk}{2\pi}
                                  \theta(\omega J_z - \mu - |k|) \biggr] \notag\\
                    &= \Bigl(\omega - 2\mu\Bigr)\frac{|eB|}{4\pi^2}
                      + \text{(orbital part)} \,,
\end{align}
and the first term is completely consistent with a combination of the
LLL relation~\eqref{eq:onedim} and the induced
density~\eqref{eq:inducedrho}.  The second term is nothing but the CSE
formula~\eqref{eq:cse}.

For higher LL contributions, we note that
$\varepsilon^{(\uparrow)}_{n,l,k}=\varepsilon^{(\downarrow)}_{n+1,l-1,k}$
and
$\bar{\varepsilon}^{(\downarrow)}_{n,l,k}
=\bar{\varepsilon}^{(\uparrow)}_{n+1,l+1,k}$.  By incrementing $n$ and
shifting $l$ accordingly, we get:
\begin{align}
  j_5^z &= \frac{1}{S_\perp}
  \sum_{n>0,l} \int \frac{dk}{2\pi} \biggl\{
  \frac{k^2}{\varepsilon^{(\downarrow)2}_{n,l,k}} \Bigl[
  \theta(\omega J_z + \omega + \mu - \varepsilon^{(\downarrow)}_{n,l,k}) \notag\\
        &\qquad - \theta(\omega J_z + \mu -
          \varepsilon^{(\downarrow)}_{n,l,k}) \Bigr]
          + \frac{k^2}{\bar{\varepsilon}^{(\uparrow)2}_{n,l,k}} \Bigl[
          \theta(\omega J_z - \mu -
          \bar{\varepsilon}^{(\uparrow)}_{n,l,k}) \notag\\
         &\qquad - \theta(\omega  J_z-\omega - \mu -
           \bar{\varepsilon}^{(\uparrow)}_{n,l,k}) \Bigr] \biggr\} \notag\\
  &= \frac{1}{S_\perp} \sum_{n>0,l} \int \frac{dk}{2\pi}
  \frac{k^2}{\varepsilon^{(\downarrow)2}_{n,l,k}} \biggl[
  \theta(\omega J_z + \omega + \mu - \varepsilon^{(\downarrow)}_{n,l,k}) \notag\\
        &\qquad - \theta(\omega J_z + \mu -
          \varepsilon^{(\downarrow)}_{n,l,k})
          + \theta(-\omega J_z - \mu -
          \varepsilon^{(\downarrow)}_{n,l,k}) \notag\\
         &\qquad - \theta(-\omega  J_z -\omega - \mu -
          \varepsilon^{(\downarrow)}_{n,l,k}) \biggr] \notag\\
  &\simeq \frac{1}{S_\perp}
  \sum_{n>0,l} \int \frac{dk}{2\pi}
  \frac{k^2 \,\omega}{\varepsilon^{(\downarrow)2}_{n,l,k}}
         \Bigl[ \delta(\mu - \varepsilon^{(\downarrow)}_{n,l,k})
          + \delta(- \mu -
          \varepsilon^{(\downarrow)}_{n,l,k}) \Bigr] \,.
\label{eq:cve_deriv}
\end{align}
Here, in the last step, we made expansion in terms of $\omega \ll 1$,
which is adopted in the standard derivation of the CVE formula.  For
$\mu>0$ the second term from the anti-particle is vanishing.  The term
involving $J_z$ is of higher order in the $\omega$ expansion.  Then,
in the limit of $eB\to 0$, we can replace
$(1/S_\perp)\sum_{n,l} \int dk_\perp^2/(2\pi)^2$ [as we noted
below Eq.~\eqref{eq:replace}].  The phase space integration leads to
\begin{equation}
  j_5^z \simeq \omega\int \frac{d^3 k}{(2\pi)^3}
  \frac{k_z^2}{\varepsilon_k^2} \delta(\mu-\varepsilon_k)
  = \frac{\mu^2 \omega}{6\pi^2} \,.
  \label{eq:cve/3}
\end{equation}
One may think that this result has a wrong coefficient by $1/3$ as
compared to the CVE formula, but the calculation is completely correct
and this factor $1/3$ problem has been already resolved well.

The remaining $2/3$ contributions are accounted for by the
magnetization current.  The magnetization current is of great
importance to maintain Lorentz
symmetry~\cite{Son:2012zy,Manuel:2013zaa,Chen:2014cla,Chen:2015gta},
which is also associated with the side-jump
effect~\cite{PhysRevB.2.4559,RevModPhys.82.1539,RevModPhys.82.1959}.
For recent discussions on the side-jump effect within the framework of
the chiral kinetic theory, see
Refs.~\cite{Hidaka:2016yjf,Hidaka:2017auj,Hidaka:2018ekt,Huang:2018aly}.

The most interesting observation in the calculation of
Eq.~\eqref{eq:cve_deriv} is that the energy derivative appears in the
small $\omega$ limit from a shift in $l$ for opposite
$(\uparrow,\downarrow)$.  This derivative is already seen in the
original field-theoretical computation~\cite{Vilenkin:1980zv}, but our
mode decomposed description gives a plain explanation of the
microscopic origin of the $\omega$ derivative.  Now, we could have
reproduced the correct CVE formula, but this requires extra
discussions on the magnetic contribution to the angular momentum.  In
fact, it would be an extremely interesting question how the CVE
formula would be modified by nonzero $e\bB$.  One might think that
Eq.~\eqref{eq:cve_deriv} already gives the answer, but even for
numerical analysis, more careful treatments (for example, imposing a
proper boundary condition at $r=R$ to discretize the transverse
momenta;  see Ref.~\cite{Chen:2017xrj}) are crucial.  Since these
vital improvements are technically involved, and since this is a
tremendously important problem on its own, we will report all such
details in another publication.

\section{Summary}

In this work we introduced the mode decomposed description of
fermionic wave-functions at finite magnetic field $\bB$.  In this
framework of the mode decomposed description fermion fields are
presented with quantum numbers corresponding to the helicity and the
orbital angular momentum $l$.  Using these wave-functions we obtained
explicit forms of the mode-by-mode contributions to the density, the
chirality density, the vector current, and the axial vector current.
We found a remarkable relation between the vector current mode
$j_{n,l,k}^{z(\uparrow,\downarrow)}$ and the chirality density
$\rho_{5\,n,l,k}^{(\uparrow,\downarrow)}$.  We would call this
surprisingly simple relation,
$j_{n,l,k}^{z(\uparrow,\downarrow)}=\pm\rho_{5\,n,l,k}^{(\uparrow,\downarrow)}$,
the mode decomposed CME{}.  We would emphasize noteworthy advantages
of the mode decomposed CME as follows:  First of all, not the chiral
chemical potential but the chirality density is directly associated
with the vector current.  In this way the mode decomposed CME can
evade the problematic chiral chemical potential which may cause
controversial interpretations.  We confirmed that an appropriate
superposition of modes weighted with the chiral chemical potential can
correctly reproduce the usual CME from the mode decomposed formula.
Second of all, the mode decomposed CME does hold for all Landau levels
and any mass.  We note that a similar relation between the axial
current and the density is derived only for the LLLs.  Such an elegant
relation reminds us that the CME is anomaly protected, i.e., the
coefficient in front of the magnetic field is unaffected by any
infrared scales.  It would be an intriguing future problem to study
how the mode decomposed CME may undergo a change by interaction
effects.  Probably only the LLL part is anomaly protected as the CME
is, and yet, the mode decomposed CME makes sense as long as the
quasi-particle picture is valid (for example, in the mean-field
approximation in which interaction effects are renormalized in a
dressed mass).

Furthermore, we applied the mode decomposed formulation to rotating
fermion systems.  It is known that a finite rotation shifts the
one-particle energy by a cranking term, $\bomega\cdot \bJ$, which has
analogy with the $\mu N$ term in finite density systems.  In our mode
decomposed formulation a mode with a fixed $l$ is specifically
considered, and our analysis, in this sense, is regarded as the
canonical approach in which $N$ for finite density is directly fixed.
The grand canonical approach fixes a finite $\mu$ for finite density
and a finite $\omega$ for finite rotation.  We note that the canonical
approach to fix $l$ and the grand canonical approach to fix $\omega$
cannot be equivalent since the causality bound prevents us from taking
the thermodynamic limit.  One application of the mode decomposed
description is the analysis on the density redistribution
$\propto \bomega\cdot\bB$ as found in Ref.~\cite{Hattori:2016njk}.
The favored LLLs at finite $\bB$ for particles and anti-particles are
directed in opposite orientations, and a finite rotation induces
imbalance between particles and anti-particles.  We also checked that
the CVE formula apart from the magnetization current results from the
weighted sum of higher Landau levels.  Non-central heavy ion collisions
produce hot and dense matter with strong magnetic field and huge
vorticity.  Our formulation can potentially have a tight connection to
heavy ion collisions at finite impact parameter.

In closing, let us mention a prospective possibility of implementing
electron vortex beams.  The mode decomposed formulation as presented
in this paper is to be translated into the theory of electron vortex
beams for which wave-functions with fixed $l$ are concerned.  Phase
vortices are receiving increased attention in particle collision
physics~\cite{Ivanov:2019vxe}.
Applying our formulation to this topic could be an intriguing possibility.
Furthermore, we would say that the idea
could be directly tested in an optical setup;  a particularly interesting
example is the mode decomposed CME manifested by the helicity projection
as discussed in this work.  If relativistic electron beams with a
certain helicity selected out are prepared, a possible realization of
the mode decomposed CME should be the generation of finite chirality
density in response to the imposed helicity projected current through
Eq.~\eqref{eq:mdcme}.  In other words, one can handle the chirality
density by manipulating the helicity projected current, and vice versa.
This situation of electron beams makes a sharp contrast to that in
heavy-ion collision experiments.  In the latter a finite chirality
density could be created only locally in each collision event, while
only averaged quantities such as fluctuations are measurable.

We would emphasize an attractive point of electron vortex beams.  Now,
nonrelativistic beams produce vortex beams with $l$ separated, and
relativistic beams are going to be available in the near future.  The
impact is immense;  it is still quite difficult to control rotation in
experiments.  The heavy-ion collision can certainly generate
relativistic swirl, but rotation itself is not really controllable.
It is even challenging to control rotation in tabletop experiments.  A
large $l$ instead of a large $\omega$ could be an alternative method
to investigate rotating fermions.  For the purpose to establish more
reliable predictions for possible experiments especially when a finite
magnetic field coexists, it is indispensable to develop a theoretical
framework in which electromagnetic angular momenta are properly taken
into account (see, for example, Ref.~\cite{Fukushima:2020qta} for
significance of electromagnetic contributions).  All these issues
including optical applications for chiral physics deserve further
investigations.

\begin{acknowledgments}
  We thank
  Xu-Guang~Huang and
  Kazuya~Mameda
  for useful discussions.
  K.~F.\ was supported by Japan Society for the Promotion
  of Science (JSPS) KAKENHI Grant Nos.\ 18H01211 and 19K21874.
\end{acknowledgments}

\bibliographystyle{apsrev4-2}
\bibliography{vortex}

\begin{thebibliography}{93}%
\makeatletter
\providecommand \@ifxundefined [1]{%
 \@ifx{#1\undefined}
}%
\providecommand \@ifnum [1]{%
 \ifnum #1\expandafter \@firstoftwo
 \else \expandafter \@secondoftwo
 \fi
}%
\providecommand \@ifx [1]{%
 \ifx #1\expandafter \@firstoftwo
 \else \expandafter \@secondoftwo
 \fi
}%
\providecommand \natexlab [1]{#1}%
\providecommand \enquote  [1]{``#1''}%
\providecommand \bibnamefont  [1]{#1}%
\providecommand \bibfnamefont [1]{#1}%
\providecommand \citenamefont [1]{#1}%
\providecommand \href@noop [0]{\@secondoftwo}%
\providecommand \href [0]{\begingroup \@sanitize@url \@href}%
\providecommand \@href[1]{\@@startlink{#1}\@@href}%
\providecommand \@@href[1]{\endgroup#1\@@endlink}%
\providecommand \@sanitize@url [0]{\catcode `\\12\catcode `\$12\catcode
  `\&12\catcode `\#12\catcode `\^12\catcode `\_12\catcode `\%12\relax}%
\providecommand \@@startlink[1]{}%
\providecommand \@@endlink[0]{}%
\providecommand \url  [0]{\begingroup\@sanitize@url \@url }%
\providecommand \@url [1]{\endgroup\@href {#1}{\urlprefix }}%
\providecommand \urlprefix  [0]{URL }%
\providecommand \Eprint [0]{\href }%
\providecommand \doibase [0]{https://doi.org/}%
\providecommand \selectlanguage [0]{\@gobble}%
\providecommand \bibinfo  [0]{\@secondoftwo}%
\providecommand \bibfield  [0]{\@secondoftwo}%
\providecommand \translation [1]{[#1]}%
\providecommand \BibitemOpen [0]{}%
\providecommand \bibitemStop [0]{}%
\providecommand \bibitemNoStop [0]{.\EOS\space}%
\providecommand \EOS [0]{\spacefactor3000\relax}%
\providecommand \BibitemShut  [1]{\csname bibitem#1\endcsname}%
\let\auto@bib@innerbib\@empty
\bibitem [{\citenamefont {Adler}(1969)}]{Adler:1969gk}%
  \BibitemOpen
  \bibfield  {author} {\bibinfo {author} {\bibfnamefont {S.~L.}\ \bibnamefont
  {Adler}},\ }\href {https://doi.org/10.1103/PhysRev.177.2426} {\bibfield
  {journal} {\bibinfo  {journal} {Phys. Rev.}\ }\textbf {\bibinfo {volume}
  {177}},\ \bibinfo {pages} {2426} (\bibinfo {year} {1969})}\BibitemShut
  {NoStop}%
\bibitem [{\citenamefont {Bell}\ and\ \citenamefont
  {Jackiw}(1969)}]{Bell:1969ts}%
  \BibitemOpen
  \bibfield  {author} {\bibinfo {author} {\bibfnamefont {J.~S.}\ \bibnamefont
  {Bell}}\ and\ \bibinfo {author} {\bibfnamefont {R.}~\bibnamefont {Jackiw}},\
  }\href {https://doi.org/10.1007/BF02823296} {\bibfield  {journal} {\bibinfo
  {journal} {Nuovo Cim. A}\ }\textbf {\bibinfo {volume} {60}},\ \bibinfo
  {pages} {47} (\bibinfo {year} {1969})}\BibitemShut {NoStop}%
\bibitem [{\citenamefont {Witten}(1979)}]{Witten:1979vv}%
  \BibitemOpen
  \bibfield  {author} {\bibinfo {author} {\bibfnamefont {E.}~\bibnamefont
  {Witten}},\ }\href {https://doi.org/10.1016/0550-3213(79)90031-2} {\bibfield
  {journal} {\bibinfo  {journal} {Nucl.\ Phys.\ B}\ }\textbf {\bibinfo {volume}
  {156}},\ \bibinfo {pages} {269} (\bibinfo {year} {1979})}\BibitemShut
  {NoStop}%
\bibitem [{\citenamefont {Veneziano}(1979)}]{Veneziano:1979ec}%
  \BibitemOpen
  \bibfield  {author} {\bibinfo {author} {\bibfnamefont {G.}~\bibnamefont
  {Veneziano}},\ }\href {https://doi.org/10.1016/0550-3213(79)90332-8}
  {\bibfield  {journal} {\bibinfo  {journal} {Nucl.\ Phys.\ B}\ }\textbf
  {\bibinfo {volume} {159}},\ \bibinfo {pages} {213} (\bibinfo {year}
  {1979})}\BibitemShut {NoStop}%
\bibitem [{\citenamefont {Kharzeev}\ \emph {et~al.}(1998)\citenamefont
  {Kharzeev}, \citenamefont {Pisarski},\ and\ \citenamefont
  {Tytgat}}]{Kharzeev:1998kz}%
  \BibitemOpen
  \bibfield  {author} {\bibinfo {author} {\bibfnamefont {D.}~\bibnamefont
  {Kharzeev}}, \bibinfo {author} {\bibfnamefont {R.}~\bibnamefont {Pisarski}},\
  and\ \bibinfo {author} {\bibfnamefont {M.~H.}\ \bibnamefont {Tytgat}},\
  }\href {https://doi.org/10.1103/PhysRevLett.81.512} {\bibfield  {journal}
  {\bibinfo  {journal} {Phys.\ Rev.\ Lett.}\ }\textbf {\bibinfo {volume}
  {81}},\ \bibinfo {pages} {512} (\bibinfo {year} {1998})},\ \Eprint
  {https://arxiv.org/abs/hep-ph/9804221} {arXiv:hep-ph/9804221} \BibitemShut
  {NoStop}%
\bibitem [{\citenamefont {Kharzeev}\ \emph {et~al.}(2008)\citenamefont
  {Kharzeev}, \citenamefont {McLerran},\ and\ \citenamefont
  {Warringa}}]{Kharzeev:2007jp}%
  \BibitemOpen
  \bibfield  {author} {\bibinfo {author} {\bibfnamefont {D.~E.}\ \bibnamefont
  {Kharzeev}}, \bibinfo {author} {\bibfnamefont {L.~D.}\ \bibnamefont
  {McLerran}},\ and\ \bibinfo {author} {\bibfnamefont {H.~J.}\ \bibnamefont
  {Warringa}},\ }\href {https://doi.org/10.1016/j.nuclphysa.2008.02.298}
  {\bibfield  {journal} {\bibinfo  {journal} {Nucl. Phys. A}\ }\textbf
  {\bibinfo {volume} {803}},\ \bibinfo {pages} {227} (\bibinfo {year}
  {2008})},\ \Eprint {https://arxiv.org/abs/0711.0950} {arXiv:0711.0950
  [hep-ph]} \BibitemShut {NoStop}%
\bibitem [{\citenamefont {Vilenkin}(1980{\natexlab{a}})}]{Vilenkin:1980fu}%
  \BibitemOpen
  \bibfield  {author} {\bibinfo {author} {\bibfnamefont {A.}~\bibnamefont
  {Vilenkin}},\ }\href {https://doi.org/10.1103/PhysRevD.22.3080} {\bibfield
  {journal} {\bibinfo  {journal} {Phys.\ Rev.\ D}\ }\textbf {\bibinfo {volume}
  {22}},\ \bibinfo {pages} {3080} (\bibinfo {year}
  {1980}{\natexlab{a}})}\BibitemShut {NoStop}%
\bibitem [{\citenamefont {Fukushima}\ \emph {et~al.}(2008)\citenamefont
  {Fukushima}, \citenamefont {Kharzeev},\ and\ \citenamefont
  {Warringa}}]{Fukushima:2008xe}%
  \BibitemOpen
  \bibfield  {author} {\bibinfo {author} {\bibfnamefont {K.}~\bibnamefont
  {Fukushima}}, \bibinfo {author} {\bibfnamefont {D.~E.}\ \bibnamefont
  {Kharzeev}},\ and\ \bibinfo {author} {\bibfnamefont {H.~J.}\ \bibnamefont
  {Warringa}},\ }\href {https://doi.org/10.1103/PhysRevD.78.074033} {\bibfield
  {journal} {\bibinfo  {journal} {Phys. Rev. D}\ }\textbf {\bibinfo {volume}
  {78}},\ \bibinfo {pages} {074033} (\bibinfo {year} {2008})},\ \Eprint
  {https://arxiv.org/abs/0808.3382} {arXiv:0808.3382 [hep-ph]} \BibitemShut
  {NoStop}%
\bibitem [{\citenamefont {Kharzeev}\ and\ \citenamefont
  {Warringa}(2009)}]{Kharzeev:2009pj}%
  \BibitemOpen
  \bibfield  {author} {\bibinfo {author} {\bibfnamefont {D.~E.}\ \bibnamefont
  {Kharzeev}}\ and\ \bibinfo {author} {\bibfnamefont {H.~J.}\ \bibnamefont
  {Warringa}},\ }\href {https://doi.org/10.1103/PhysRevD.80.034028} {\bibfield
  {journal} {\bibinfo  {journal} {Phys.\ Rev.\ D}\ }\textbf {\bibinfo {volume}
  {80}},\ \bibinfo {pages} {034028} (\bibinfo {year} {2009})},\ \Eprint
  {https://arxiv.org/abs/0907.5007} {arXiv:0907.5007 [hep-ph]} \BibitemShut
  {NoStop}%
\bibitem [{\citenamefont {Abelev}\ \emph {et~al.}(2009)\citenamefont {Abelev}
  \emph {et~al.}}]{Abelev:2009ac}%
  \BibitemOpen
  \bibfield  {author} {\bibinfo {author} {\bibfnamefont {B.~I.}\ \bibnamefont
  {Abelev}} \emph {et~al.} (\bibinfo {collaboration} {STAR}),\ }\href
  {https://doi.org/10.1103/PhysRevLett.103.251601} {\bibfield  {journal}
  {\bibinfo  {journal} {Phys. Rev. Lett.}\ }\textbf {\bibinfo {volume} {103}},\
  \bibinfo {pages} {251601} (\bibinfo {year} {2009})},\ \Eprint
  {https://arxiv.org/abs/0909.1739} {arXiv:0909.1739 [nucl-ex]} \BibitemShut
  {NoStop}%
\bibitem [{\citenamefont {Abelev}\ \emph {et~al.}(2010)\citenamefont {Abelev}
  \emph {et~al.}}]{Abelev:2009ad}%
  \BibitemOpen
  \bibfield  {author} {\bibinfo {author} {\bibfnamefont {B.}~\bibnamefont
  {Abelev}} \emph {et~al.} (\bibinfo {collaboration} {STAR}),\ }\href
  {https://doi.org/10.1103/PhysRevC.81.054908} {\bibfield  {journal} {\bibinfo
  {journal} {Phys.\ Rev.\ C}\ }\textbf {\bibinfo {volume} {81}},\ \bibinfo
  {pages} {054908} (\bibinfo {year} {2010})},\ \Eprint
  {https://arxiv.org/abs/0909.1717} {arXiv:0909.1717 [nucl-ex]} \BibitemShut
  {NoStop}%
\bibitem [{\citenamefont {Kharzeev}\ and\ \citenamefont
  {Son}(2011)}]{Kharzeev:2010gr}%
  \BibitemOpen
  \bibfield  {author} {\bibinfo {author} {\bibfnamefont {D.~E.}\ \bibnamefont
  {Kharzeev}}\ and\ \bibinfo {author} {\bibfnamefont {D.~T.}\ \bibnamefont
  {Son}},\ }\href {https://doi.org/10.1103/PhysRevLett.106.062301} {\bibfield
  {journal} {\bibinfo  {journal} {Phys.\ Rev.\ Lett.}\ }\textbf {\bibinfo
  {volume} {106}},\ \bibinfo {pages} {062301} (\bibinfo {year} {2011})},\
  \Eprint {https://arxiv.org/abs/1010.0038} {arXiv:1010.0038 [hep-ph]}
  \BibitemShut {NoStop}%
\bibitem [{\citenamefont {Abelev}\ \emph {et~al.}(2013)\citenamefont {Abelev}
  \emph {et~al.}}]{Abelev:2012pa}%
  \BibitemOpen
  \bibfield  {author} {\bibinfo {author} {\bibfnamefont {B.}~\bibnamefont
  {Abelev}} \emph {et~al.} (\bibinfo {collaboration} {ALICE}),\ }\href
  {https://doi.org/10.1103/PhysRevLett.110.012301} {\bibfield  {journal}
  {\bibinfo  {journal} {Phys. Rev. Lett.}\ }\textbf {\bibinfo {volume} {110}},\
  \bibinfo {pages} {012301} (\bibinfo {year} {2013})},\ \Eprint
  {https://arxiv.org/abs/1207.0900} {arXiv:1207.0900 [nucl-ex]} \BibitemShut
  {NoStop}%
\bibitem [{\citenamefont {Huang}(2016)}]{Huang:2015oca}%
  \BibitemOpen
  \bibfield  {author} {\bibinfo {author} {\bibfnamefont {X.-G.}\ \bibnamefont
  {Huang}},\ }\href {https://doi.org/10.1088/0034-4885/79/7/076302} {\bibfield
  {journal} {\bibinfo  {journal} {Rept. Prog. Phys.}\ }\textbf {\bibinfo
  {volume} {79}},\ \bibinfo {pages} {076302} (\bibinfo {year} {2016})},\
  \Eprint {https://arxiv.org/abs/1509.04073} {arXiv:1509.04073 [nucl-th]}
  \BibitemShut {NoStop}%
\bibitem [{\citenamefont {Kharzeev}\ \emph {et~al.}(2016)\citenamefont
  {Kharzeev}, \citenamefont {Liao}, \citenamefont {Voloshin},\ and\
  \citenamefont {Wang}}]{Kharzeev:2015znc}%
  \BibitemOpen
  \bibfield  {author} {\bibinfo {author} {\bibfnamefont {D.~E.}\ \bibnamefont
  {Kharzeev}}, \bibinfo {author} {\bibfnamefont {J.}~\bibnamefont {Liao}},
  \bibinfo {author} {\bibfnamefont {S.~A.}\ \bibnamefont {Voloshin}},\ and\
  \bibinfo {author} {\bibfnamefont {G.}~\bibnamefont {Wang}},\ }\href
  {https://doi.org/10.1016/j.ppnp.2016.01.001} {\bibfield  {journal} {\bibinfo
  {journal} {Prog. Part. Nucl. Phys.}\ }\textbf {\bibinfo {volume} {88}},\
  \bibinfo {pages} {1} (\bibinfo {year} {2016})},\ \Eprint
  {https://arxiv.org/abs/1511.04050} {arXiv:1511.04050 [hep-ph]} \BibitemShut
  {NoStop}%
\bibitem [{\citenamefont {Son}\ and\ \citenamefont
  {Spivak}(2013)}]{Son:2012bg}%
  \BibitemOpen
  \bibfield  {author} {\bibinfo {author} {\bibfnamefont {D.}~\bibnamefont
  {Son}}\ and\ \bibinfo {author} {\bibfnamefont {B.}~\bibnamefont {Spivak}},\
  }\href {https://doi.org/10.1103/PhysRevB.88.104412} {\bibfield  {journal}
  {\bibinfo  {journal} {Phys.\ Rev.\ B}\ }\textbf {\bibinfo {volume} {88}},\
  \bibinfo {pages} {104412} (\bibinfo {year} {2013})},\ \Eprint
  {https://arxiv.org/abs/1206.1627} {arXiv:1206.1627 [cond-mat.mes-hall]}
  \BibitemShut {NoStop}%
\bibitem [{\citenamefont {Li}\ \emph {et~al.}(2016)\citenamefont {Li},
  \citenamefont {Kharzeev}, \citenamefont {Zhang}, \citenamefont {Huang},
  \citenamefont {Pletikosic}, \citenamefont {Fedorov}, \citenamefont {Zhong},
  \citenamefont {Schneeloch}, \citenamefont {Gu},\ and\ \citenamefont
  {Valla}}]{Li:2014bha}%
  \BibitemOpen
  \bibfield  {author} {\bibinfo {author} {\bibfnamefont {Q.}~\bibnamefont
  {Li}}, \bibinfo {author} {\bibfnamefont {D.~E.}\ \bibnamefont {Kharzeev}},
  \bibinfo {author} {\bibfnamefont {C.}~\bibnamefont {Zhang}}, \bibinfo
  {author} {\bibfnamefont {Y.}~\bibnamefont {Huang}}, \bibinfo {author}
  {\bibfnamefont {I.}~\bibnamefont {Pletikosic}}, \bibinfo {author}
  {\bibfnamefont {A.~V.}\ \bibnamefont {Fedorov}}, \bibinfo {author}
  {\bibfnamefont {R.~D.}\ \bibnamefont {Zhong}}, \bibinfo {author}
  {\bibfnamefont {J.~A.}\ \bibnamefont {Schneeloch}}, \bibinfo {author}
  {\bibfnamefont {G.~D.}\ \bibnamefont {Gu}},\ and\ \bibinfo {author}
  {\bibfnamefont {T.}~\bibnamefont {Valla}},\ }\href
  {https://doi.org/10.1038/nphys3648} {\bibfield  {journal} {\bibinfo
  {journal} {Nature Phys.}\ }\textbf {\bibinfo {volume} {12}},\ \bibinfo
  {pages} {550} (\bibinfo {year} {2016})},\ \Eprint
  {https://arxiv.org/abs/1412.6543} {arXiv:1412.6543 [cond-mat.str-el]}
  \BibitemShut {NoStop}%
\bibitem [{\citenamefont {Huang}\ \emph {et~al.}(2015)\citenamefont {Huang}
  \emph {et~al.}}]{Huang:2015eia}%
  \BibitemOpen
  \bibfield  {author} {\bibinfo {author} {\bibfnamefont {X.}~\bibnamefont
  {Huang}} \emph {et~al.},\ }\href {https://doi.org/10.1103/PhysRevX.5.031023}
  {\bibfield  {journal} {\bibinfo  {journal} {Phys. Rev. X}\ }\textbf {\bibinfo
  {volume} {5}},\ \bibinfo {pages} {031023} (\bibinfo {year} {2015})},\ \Eprint
  {https://arxiv.org/abs/1503.01304} {arXiv:1503.01304 [cond-mat.mtrl-sci]}
  \BibitemShut {NoStop}%
\bibitem [{\citenamefont {Li}\ and\ \citenamefont
  {Kharzeev}(2016)}]{Li:2016vlc}%
  \BibitemOpen
  \bibfield  {author} {\bibinfo {author} {\bibfnamefont {Q.}~\bibnamefont
  {Li}}\ and\ \bibinfo {author} {\bibfnamefont {D.~E.}\ \bibnamefont
  {Kharzeev}},\ }\href {https://doi.org/10.1016/j.nuclphysa.2016.03.055}
  {\bibfield  {journal} {\bibinfo  {journal} {Nucl. Phys. A}\ }\textbf
  {\bibinfo {volume} {956}},\ \bibinfo {pages} {107} (\bibinfo {year}
  {2016})}\BibitemShut {NoStop}%
\bibitem [{\citenamefont {Copinger}\ \emph {et~al.}(2018)\citenamefont
  {Copinger}, \citenamefont {Fukushima},\ and\ \citenamefont
  {Pu}}]{Copinger:2018ftr}%
  \BibitemOpen
  \bibfield  {author} {\bibinfo {author} {\bibfnamefont {P.}~\bibnamefont
  {Copinger}}, \bibinfo {author} {\bibfnamefont {K.}~\bibnamefont
  {Fukushima}},\ and\ \bibinfo {author} {\bibfnamefont {S.}~\bibnamefont
  {Pu}},\ }\href {https://doi.org/10.1103/PhysRevLett.121.261602} {\bibfield
  {journal} {\bibinfo  {journal} {Phys.\ Rev.\ Lett.}\ }\textbf {\bibinfo
  {volume} {121}},\ \bibinfo {pages} {261602} (\bibinfo {year} {2018})},\
  \Eprint {https://arxiv.org/abs/1807.04416} {arXiv:1807.04416 [hep-th]}
  \BibitemShut {NoStop}%
\bibitem [{\citenamefont {Son}\ and\ \citenamefont
  {Zhitnitsky}(2004)}]{Son:2004tq}%
  \BibitemOpen
  \bibfield  {author} {\bibinfo {author} {\bibfnamefont {D.~T.}\ \bibnamefont
  {Son}}\ and\ \bibinfo {author} {\bibfnamefont {A.~R.}\ \bibnamefont
  {Zhitnitsky}},\ }\href {https://doi.org/10.1103/PhysRevD.70.074018}
  {\bibfield  {journal} {\bibinfo  {journal} {Phys. Rev. D}\ }\textbf {\bibinfo
  {volume} {70}},\ \bibinfo {pages} {074018} (\bibinfo {year} {2004})},\
  \Eprint {https://arxiv.org/abs/hep-ph/0405216} {arXiv:hep-ph/0405216
  [hep-ph]} \BibitemShut {NoStop}%
\bibitem [{\citenamefont {Metlitski}\ and\ \citenamefont
  {Zhitnitsky}(2005)}]{Metlitski:2005pr}%
  \BibitemOpen
  \bibfield  {author} {\bibinfo {author} {\bibfnamefont {M.~A.}\ \bibnamefont
  {Metlitski}}\ and\ \bibinfo {author} {\bibfnamefont {A.~R.}\ \bibnamefont
  {Zhitnitsky}},\ }\href {https://doi.org/10.1103/PhysRevD.72.045011}
  {\bibfield  {journal} {\bibinfo  {journal} {Phys. Rev. D}\ }\textbf {\bibinfo
  {volume} {72}},\ \bibinfo {pages} {045011} (\bibinfo {year} {2005})},\
  \Eprint {https://arxiv.org/abs/hep-ph/0505072} {arXiv:hep-ph/0505072
  [hep-ph]} \BibitemShut {NoStop}%
\bibitem [{\citenamefont {Newman}\ and\ \citenamefont
  {Son}(2006)}]{Newman:2005as}%
  \BibitemOpen
  \bibfield  {author} {\bibinfo {author} {\bibfnamefont {G.}~\bibnamefont
  {Newman}}\ and\ \bibinfo {author} {\bibfnamefont {D.}~\bibnamefont {Son}},\
  }\href {https://doi.org/10.1103/PhysRevD.73.045006} {\bibfield  {journal}
  {\bibinfo  {journal} {Phys.\ Rev.\ D}\ }\textbf {\bibinfo {volume} {73}},\
  \bibinfo {pages} {045006} (\bibinfo {year} {2006})},\ \Eprint
  {https://arxiv.org/abs/hep-ph/0510049} {arXiv:hep-ph/0510049} \BibitemShut
  {NoStop}%
\bibitem [{\citenamefont {Vilenkin}(1979)}]{Vilenkin:1979ui}%
  \BibitemOpen
  \bibfield  {author} {\bibinfo {author} {\bibfnamefont {A.}~\bibnamefont
  {Vilenkin}},\ }\href {https://doi.org/10.1103/PhysRevD.20.1807} {\bibfield
  {journal} {\bibinfo  {journal} {Phys.\ Rev.\ D}\ }\textbf {\bibinfo {volume}
  {20}},\ \bibinfo {pages} {1807} (\bibinfo {year} {1979})}\BibitemShut
  {NoStop}%
\bibitem [{\citenamefont {Vilenkin}(1980{\natexlab{b}})}]{Vilenkin:1980zv}%
  \BibitemOpen
  \bibfield  {author} {\bibinfo {author} {\bibfnamefont {A.}~\bibnamefont
  {Vilenkin}},\ }\href {https://doi.org/10.1103/PhysRevD.21.2260} {\bibfield
  {journal} {\bibinfo  {journal} {Phys. Rev. D}\ }\textbf {\bibinfo {volume}
  {21}},\ \bibinfo {pages} {2260} (\bibinfo {year}
  {1980}{\natexlab{b}})}\BibitemShut {NoStop}%
\bibitem [{\citenamefont {Erdmenger}\ \emph {et~al.}(2009)\citenamefont
  {Erdmenger}, \citenamefont {Haack}, \citenamefont {Kaminski},\ and\
  \citenamefont {Yarom}}]{Erdmenger:2008rm}%
  \BibitemOpen
  \bibfield  {author} {\bibinfo {author} {\bibfnamefont {J.}~\bibnamefont
  {Erdmenger}}, \bibinfo {author} {\bibfnamefont {M.}~\bibnamefont {Haack}},
  \bibinfo {author} {\bibfnamefont {M.}~\bibnamefont {Kaminski}},\ and\
  \bibinfo {author} {\bibfnamefont {A.}~\bibnamefont {Yarom}},\ }\href
  {https://doi.org/10.1088/1126-6708/2009/01/055} {\bibfield  {journal}
  {\bibinfo  {journal} {JHEP}\ }\textbf {\bibinfo {volume} {01}},\ \bibinfo
  {pages} {055}},\ \Eprint {https://arxiv.org/abs/0809.2488} {arXiv:0809.2488
  [hep-th]} \BibitemShut {NoStop}%
\bibitem [{\citenamefont {Son}\ and\ \citenamefont
  {Surowka}(2009)}]{Son:2009tf}%
  \BibitemOpen
  \bibfield  {author} {\bibinfo {author} {\bibfnamefont {D.~T.}\ \bibnamefont
  {Son}}\ and\ \bibinfo {author} {\bibfnamefont {P.}~\bibnamefont {Surowka}},\
  }\href {https://doi.org/10.1103/PhysRevLett.103.191601} {\bibfield  {journal}
  {\bibinfo  {journal} {Phys. Rev. Lett.}\ }\textbf {\bibinfo {volume} {103}},\
  \bibinfo {pages} {191601} (\bibinfo {year} {2009})},\ \Eprint
  {https://arxiv.org/abs/0906.5044} {arXiv:0906.5044 [hep-th]} \BibitemShut
  {NoStop}%
\bibitem [{\citenamefont {Landsteiner}\ \emph {et~al.}(2011)\citenamefont
  {Landsteiner}, \citenamefont {Megias},\ and\ \citenamefont
  {Pena-Benitez}}]{Landsteiner:2011cp}%
  \BibitemOpen
  \bibfield  {author} {\bibinfo {author} {\bibfnamefont {K.}~\bibnamefont
  {Landsteiner}}, \bibinfo {author} {\bibfnamefont {E.}~\bibnamefont
  {Megias}},\ and\ \bibinfo {author} {\bibfnamefont {F.}~\bibnamefont
  {Pena-Benitez}},\ }\href {https://doi.org/10.1103/PhysRevLett.107.021601}
  {\bibfield  {journal} {\bibinfo  {journal} {Phys.\ Rev.\ Lett.}\ }\textbf
  {\bibinfo {volume} {107}},\ \bibinfo {pages} {021601} (\bibinfo {year}
  {2011})},\ \Eprint {https://arxiv.org/abs/1103.5006} {arXiv:1103.5006
  [hep-ph]} \BibitemShut {NoStop}%
\bibitem [{\citenamefont {Flachi}\ and\ \citenamefont
  {Fukushima}(2018)}]{Flachi:2017vlp}%
  \BibitemOpen
  \bibfield  {author} {\bibinfo {author} {\bibfnamefont {A.}~\bibnamefont
  {Flachi}}\ and\ \bibinfo {author} {\bibfnamefont {K.}~\bibnamefont
  {Fukushima}},\ }\href {https://doi.org/10.1103/PhysRevD.98.096011} {\bibfield
   {journal} {\bibinfo  {journal} {Phys. Rev. D}\ }\textbf {\bibinfo {volume}
  {98}},\ \bibinfo {pages} {096011} (\bibinfo {year} {2018})},\ \Eprint
  {https://arxiv.org/abs/1702.04753} {arXiv:1702.04753 [hep-th]} \BibitemShut
  {NoStop}%
\bibitem [{\citenamefont {Abramchuk}\ \emph {et~al.}(2018)\citenamefont
  {Abramchuk}, \citenamefont {Khaidukov},\ and\ \citenamefont
  {Zubkov}}]{Abramchuk:2018jhd}%
  \BibitemOpen
  \bibfield  {author} {\bibinfo {author} {\bibfnamefont {R.}~\bibnamefont
  {Abramchuk}}, \bibinfo {author} {\bibfnamefont {Z.~V.}\ \bibnamefont
  {Khaidukov}},\ and\ \bibinfo {author} {\bibfnamefont {M.~A.}\ \bibnamefont
  {Zubkov}},\ }\href {https://doi.org/10.1103/PhysRevD.98.076013} {\bibfield
  {journal} {\bibinfo  {journal} {Phys. Rev. D}\ }\textbf {\bibinfo {volume}
  {98}},\ \bibinfo {pages} {076013} (\bibinfo {year} {2018})},\ \Eprint
  {https://arxiv.org/abs/1806.02605} {arXiv:1806.02605 [hep-ph]} \BibitemShut
  {NoStop}%
\bibitem [{\citenamefont {Kharzeev}\ and\ \citenamefont
  {Zhitnitsky}(2007)}]{Kharzeev:2007tn}%
  \BibitemOpen
  \bibfield  {author} {\bibinfo {author} {\bibfnamefont {D.}~\bibnamefont
  {Kharzeev}}\ and\ \bibinfo {author} {\bibfnamefont {A.}~\bibnamefont
  {Zhitnitsky}},\ }\href {https://doi.org/10.1016/j.nuclphysa.2007.10.001}
  {\bibfield  {journal} {\bibinfo  {journal} {Nucl. Phys. A}\ }\textbf
  {\bibinfo {volume} {797}},\ \bibinfo {pages} {67} (\bibinfo {year} {2007})},\
  \Eprint {https://arxiv.org/abs/0706.1026} {arXiv:0706.1026 [hep-ph]}
  \BibitemShut {NoStop}%
\bibitem [{\citenamefont {Stephanov}\ and\ \citenamefont
  {Yin}(2012)}]{Stephanov:2012ki}%
  \BibitemOpen
  \bibfield  {author} {\bibinfo {author} {\bibfnamefont {M.}~\bibnamefont
  {Stephanov}}\ and\ \bibinfo {author} {\bibfnamefont {Y.}~\bibnamefont
  {Yin}},\ }\href {https://doi.org/10.1103/PhysRevLett.109.162001} {\bibfield
  {journal} {\bibinfo  {journal} {Phys.\ Rev.\ Lett.}\ }\textbf {\bibinfo
  {volume} {109}},\ \bibinfo {pages} {162001} (\bibinfo {year} {2012})},\
  \Eprint {https://arxiv.org/abs/1207.0747} {arXiv:1207.0747 [hep-th]}
  \BibitemShut {NoStop}%
\bibitem [{\citenamefont {McInnes}(2016)}]{McInnes:2016dwk}%
  \BibitemOpen
  \bibfield  {author} {\bibinfo {author} {\bibfnamefont {B.}~\bibnamefont
  {McInnes}},\ }\href {https://doi.org/10.1016/j.nuclphysb.2016.08.001}
  {\bibfield  {journal} {\bibinfo  {journal} {Nucl. Phys. B}\ }\textbf
  {\bibinfo {volume} {911}},\ \bibinfo {pages} {173} (\bibinfo {year}
  {2016})},\ \Eprint {https://arxiv.org/abs/1604.03669} {arXiv:1604.03669
  [hep-th]} \BibitemShut {NoStop}%
\bibitem [{\citenamefont {Gorbar}\ \emph {et~al.}(2013)\citenamefont {Gorbar},
  \citenamefont {Miransky}, \citenamefont {Shovkovy},\ and\ \citenamefont
  {Wang}}]{Gorbar:2013upa}%
  \BibitemOpen
  \bibfield  {author} {\bibinfo {author} {\bibfnamefont {E.~V.}\ \bibnamefont
  {Gorbar}}, \bibinfo {author} {\bibfnamefont {V.~A.}\ \bibnamefont
  {Miransky}}, \bibinfo {author} {\bibfnamefont {I.~A.}\ \bibnamefont
  {Shovkovy}},\ and\ \bibinfo {author} {\bibfnamefont {X.}~\bibnamefont
  {Wang}},\ }\href {https://doi.org/10.1103/PhysRevD.88.025025} {\bibfield
  {journal} {\bibinfo  {journal} {Phys. Rev. D}\ }\textbf {\bibinfo {volume}
  {88}},\ \bibinfo {pages} {025025} (\bibinfo {year} {2013})},\ \Eprint
  {https://arxiv.org/abs/1304.4606} {arXiv:1304.4606 [hep-ph]} \BibitemShut
  {NoStop}%
\bibitem [{\citenamefont {Guo}\ and\ \citenamefont {Lin}(2017)}]{Guo:2016dnm}%
  \BibitemOpen
  \bibfield  {author} {\bibinfo {author} {\bibfnamefont {E.-d.}\ \bibnamefont
  {Guo}}\ and\ \bibinfo {author} {\bibfnamefont {S.}~\bibnamefont {Lin}},\
  }\href {https://doi.org/10.1007/JHEP01(2017)111} {\bibfield  {journal}
  {\bibinfo  {journal} {JHEP}\ }\textbf {\bibinfo {volume} {01}},\ \bibinfo
  {pages} {111}},\ \Eprint {https://arxiv.org/abs/1610.05886} {arXiv:1610.05886
  [hep-th]} \BibitemShut {NoStop}%
\bibitem [{\citenamefont {Lin}\ and\ \citenamefont {Yang}(2018)}]{Lin:2018aon}%
  \BibitemOpen
  \bibfield  {author} {\bibinfo {author} {\bibfnamefont {S.}~\bibnamefont
  {Lin}}\ and\ \bibinfo {author} {\bibfnamefont {L.}~\bibnamefont {Yang}},\
  }\href {https://doi.org/10.1103/PhysRevD.98.114022} {\bibfield  {journal}
  {\bibinfo  {journal} {Phys. Rev. D}\ }\textbf {\bibinfo {volume} {98}},\
  \bibinfo {pages} {114022} (\bibinfo {year} {2018})},\ \Eprint
  {https://arxiv.org/abs/1810.02979} {arXiv:1810.02979 [nucl-th]} \BibitemShut
  {NoStop}%
\bibitem [{\citenamefont {Wang}\ \emph
  {et~al.}(2019{\natexlab{a}})\citenamefont {Wang}, \citenamefont {Guo},
  \citenamefont {Shi},\ and\ \citenamefont {Zhuang}}]{Wang:2019moi}%
  \BibitemOpen
  \bibfield  {author} {\bibinfo {author} {\bibfnamefont {Z.}~\bibnamefont
  {Wang}}, \bibinfo {author} {\bibfnamefont {X.}~\bibnamefont {Guo}}, \bibinfo
  {author} {\bibfnamefont {S.}~\bibnamefont {Shi}},\ and\ \bibinfo {author}
  {\bibfnamefont {P.}~\bibnamefont {Zhuang}},\ }\href
  {https://doi.org/10.1103/PhysRevD.100.014015} {\bibfield  {journal} {\bibinfo
   {journal} {Phys. Rev. D}\ }\textbf {\bibinfo {volume} {100}},\ \bibinfo
  {pages} {014015} (\bibinfo {year} {2019}{\natexlab{a}})},\ \Eprint
  {https://arxiv.org/abs/1903.03461} {arXiv:1903.03461 [hep-ph]} \BibitemShut
  {NoStop}%
\bibitem [{\citenamefont {Fukushima}(2019)}]{Fukushima:2018grm}%
  \BibitemOpen
  \bibfield  {author} {\bibinfo {author} {\bibfnamefont {K.}~\bibnamefont
  {Fukushima}},\ }\href {https://doi.org/10.1016/j.ppnp.2019.04.001} {\bibfield
   {journal} {\bibinfo  {journal} {Prog. Part. Nucl. Phys.}\ }\textbf {\bibinfo
  {volume} {107}},\ \bibinfo {pages} {167} (\bibinfo {year} {2019})},\ \Eprint
  {https://arxiv.org/abs/1812.08886} {arXiv:1812.08886 [hep-ph]} \BibitemShut
  {NoStop}%
\bibitem [{\citenamefont {Tuchin}(2013)}]{Tuchin:2013ie}%
  \BibitemOpen
  \bibfield  {author} {\bibinfo {author} {\bibfnamefont {K.}~\bibnamefont
  {Tuchin}},\ }\href {https://doi.org/10.1155/2013/490495} {\bibfield
  {journal} {\bibinfo  {journal} {Adv. High Energy Phys.}\ }\textbf {\bibinfo
  {volume} {2013}},\ \bibinfo {pages} {490495} (\bibinfo {year} {2013})},\
  \Eprint {https://arxiv.org/abs/1301.0099} {arXiv:1301.0099 [hep-ph]}
  \BibitemShut {NoStop}%
\bibitem [{\citenamefont {Deng}\ and\ \citenamefont
  {Huang}(2016)}]{Deng:2016gyh}%
  \BibitemOpen
  \bibfield  {author} {\bibinfo {author} {\bibfnamefont {W.-T.}\ \bibnamefont
  {Deng}}\ and\ \bibinfo {author} {\bibfnamefont {X.-G.}\ \bibnamefont
  {Huang}},\ }\href {https://doi.org/10.1103/PhysRevC.93.064907} {\bibfield
  {journal} {\bibinfo  {journal} {Phys.\ Rev.\ C}\ }\textbf {\bibinfo {volume}
  {93}},\ \bibinfo {pages} {064907} (\bibinfo {year} {2016})},\ \Eprint
  {https://arxiv.org/abs/1603.06117} {arXiv:1603.06117 [nucl-th]} \BibitemShut
  {NoStop}%
\bibitem [{\citenamefont {Adamczyk}\ \emph {et~al.}(2017)\citenamefont
  {Adamczyk} \emph {et~al.}}]{STAR:2017ckg}%
  \BibitemOpen
  \bibfield  {author} {\bibinfo {author} {\bibfnamefont {L.}~\bibnamefont
  {Adamczyk}} \emph {et~al.} (\bibinfo {collaboration} {STAR}),\ }\href
  {https://doi.org/10.1038/nature23004} {\bibfield  {journal} {\bibinfo
  {journal} {Nature}\ }\textbf {\bibinfo {volume} {548}},\ \bibinfo {pages}
  {62} (\bibinfo {year} {2017})},\ \Eprint {https://arxiv.org/abs/1701.06657}
  {arXiv:1701.06657 [nucl-ex]} \BibitemShut {NoStop}%
\bibitem [{\citenamefont {Son}\ and\ \citenamefont
  {Yamamoto}(2013)}]{Son:2012zy}%
  \BibitemOpen
  \bibfield  {author} {\bibinfo {author} {\bibfnamefont {D.~T.}\ \bibnamefont
  {Son}}\ and\ \bibinfo {author} {\bibfnamefont {N.}~\bibnamefont {Yamamoto}},\
  }\href {https://doi.org/10.1103/PhysRevD.87.085016} {\bibfield  {journal}
  {\bibinfo  {journal} {Phys. Rev. D}\ }\textbf {\bibinfo {volume} {87}},\
  \bibinfo {pages} {085016} (\bibinfo {year} {2013})},\ \Eprint
  {https://arxiv.org/abs/1210.8158} {arXiv:1210.8158 [hep-th]} \BibitemShut
  {NoStop}%
\bibitem [{\citenamefont {Hattori}\ \emph {et~al.}(2019)\citenamefont
  {Hattori}, \citenamefont {Hidaka},\ and\ \citenamefont
  {Yang}}]{Hattori:2019ahi}%
  \BibitemOpen
  \bibfield  {author} {\bibinfo {author} {\bibfnamefont {K.}~\bibnamefont
  {Hattori}}, \bibinfo {author} {\bibfnamefont {Y.}~\bibnamefont {Hidaka}},\
  and\ \bibinfo {author} {\bibfnamefont {D.-L.}\ \bibnamefont {Yang}},\ }\href
  {https://doi.org/10.1103/PhysRevD.100.096011} {\bibfield  {journal} {\bibinfo
   {journal} {Phys. Rev. D}\ }\textbf {\bibinfo {volume} {100}},\ \bibinfo
  {pages} {096011} (\bibinfo {year} {2019})},\ \Eprint
  {https://arxiv.org/abs/1903.01653} {arXiv:1903.01653 [hep-ph]} \BibitemShut
  {NoStop}%
\bibitem [{\citenamefont {Ebihara}\ \emph {et~al.}(2017)\citenamefont
  {Ebihara}, \citenamefont {Fukushima},\ and\ \citenamefont
  {Mameda}}]{Ebihara:2016fwa}%
  \BibitemOpen
  \bibfield  {author} {\bibinfo {author} {\bibfnamefont {S.}~\bibnamefont
  {Ebihara}}, \bibinfo {author} {\bibfnamefont {K.}~\bibnamefont {Fukushima}},\
  and\ \bibinfo {author} {\bibfnamefont {K.}~\bibnamefont {Mameda}},\ }\href
  {https://doi.org/10.1016/j.physletb.2016.11.010} {\bibfield  {journal}
  {\bibinfo  {journal} {Phys. Lett. B}\ }\textbf {\bibinfo {volume} {764}},\
  \bibinfo {pages} {94} (\bibinfo {year} {2017})},\ \Eprint
  {https://arxiv.org/abs/1608.00336} {arXiv:1608.00336 [hep-ph]} \BibitemShut
  {NoStop}%
\bibitem [{\citenamefont {Hattori}\ and\ \citenamefont
  {Yin}(2016)}]{Hattori:2016njk}%
  \BibitemOpen
  \bibfield  {author} {\bibinfo {author} {\bibfnamefont {K.}~\bibnamefont
  {Hattori}}\ and\ \bibinfo {author} {\bibfnamefont {Y.}~\bibnamefont {Yin}},\
  }\href {https://doi.org/10.1103/PhysRevLett.117.152002} {\bibfield  {journal}
  {\bibinfo  {journal} {Phys. Rev. Lett.}\ }\textbf {\bibinfo {volume} {117}},\
  \bibinfo {pages} {152002} (\bibinfo {year} {2016})},\ \Eprint
  {https://arxiv.org/abs/1607.01513} {arXiv:1607.01513 [hep-th]} \BibitemShut
  {NoStop}%
\bibitem [{\citenamefont {Chen}\ \emph {et~al.}(2017)\citenamefont {Chen},
  \citenamefont {Fukushima}, \citenamefont {Huang},\ and\ \citenamefont
  {Mameda}}]{Chen:2017xrj}%
  \BibitemOpen
  \bibfield  {author} {\bibinfo {author} {\bibfnamefont {H.-L.}\ \bibnamefont
  {Chen}}, \bibinfo {author} {\bibfnamefont {K.}~\bibnamefont {Fukushima}},
  \bibinfo {author} {\bibfnamefont {X.-G.}\ \bibnamefont {Huang}},\ and\
  \bibinfo {author} {\bibfnamefont {K.}~\bibnamefont {Mameda}},\ }\href
  {https://doi.org/10.1103/PhysRevD.96.054032} {\bibfield  {journal} {\bibinfo
  {journal} {Phys. Rev. D}\ }\textbf {\bibinfo {volume} {96}},\ \bibinfo
  {pages} {054032} (\bibinfo {year} {2017})},\ \Eprint
  {https://arxiv.org/abs/1707.09130} {arXiv:1707.09130 [hep-ph]} \BibitemShut
  {NoStop}%
\bibitem [{\citenamefont {Chernodub}\ and\ \citenamefont
  {Gongyo}(2017)}]{Chernodub:2017mvp}%
  \BibitemOpen
  \bibfield  {author} {\bibinfo {author} {\bibfnamefont {M.~N.}\ \bibnamefont
  {Chernodub}}\ and\ \bibinfo {author} {\bibfnamefont {S.}~\bibnamefont
  {Gongyo}},\ }\href {https://doi.org/10.1103/PhysRevD.96.096014} {\bibfield
  {journal} {\bibinfo  {journal} {Phys. Rev. D}\ }\textbf {\bibinfo {volume}
  {96}},\ \bibinfo {pages} {096014} (\bibinfo {year} {2017})},\ \Eprint
  {https://arxiv.org/abs/1706.08448} {arXiv:1706.08448 [hep-th]} \BibitemShut
  {NoStop}%
\bibitem [{\citenamefont {Liu}\ and\ \citenamefont
  {Zahed}(2018{\natexlab{a}})}]{Liu:2017zhl}%
  \BibitemOpen
  \bibfield  {author} {\bibinfo {author} {\bibfnamefont {Y.}~\bibnamefont
  {Liu}}\ and\ \bibinfo {author} {\bibfnamefont {I.}~\bibnamefont {Zahed}},\
  }\href {https://doi.org/10.1103/PhysRevD.98.014017} {\bibfield  {journal}
  {\bibinfo  {journal} {Phys. Rev. D}\ }\textbf {\bibinfo {volume} {98}},\
  \bibinfo {pages} {014017} (\bibinfo {year} {2018}{\natexlab{a}})},\ \Eprint
  {https://arxiv.org/abs/1710.02895} {arXiv:1710.02895 [hep-ph]} \BibitemShut
  {NoStop}%
\bibitem [{\citenamefont {Liu}\ and\ \citenamefont
  {Zahed}(2018{\natexlab{b}})}]{Liu:2017spl}%
  \BibitemOpen
  \bibfield  {author} {\bibinfo {author} {\bibfnamefont {Y.}~\bibnamefont
  {Liu}}\ and\ \bibinfo {author} {\bibfnamefont {I.}~\bibnamefont {Zahed}},\
  }\href {https://doi.org/10.1103/PhysRevLett.120.032001} {\bibfield  {journal}
  {\bibinfo  {journal} {Phys. Rev. Lett.}\ }\textbf {\bibinfo {volume} {120}},\
  \bibinfo {pages} {032001} (\bibinfo {year} {2018}{\natexlab{b}})},\ \Eprint
  {https://arxiv.org/abs/1711.08354} {arXiv:1711.08354 [hep-ph]} \BibitemShut
  {NoStop}%
\bibitem [{\citenamefont {Wang}\ and\ \citenamefont
  {Cao}(2018)}]{Wang:2017pje}%
  \BibitemOpen
  \bibfield  {author} {\bibinfo {author} {\bibfnamefont {L.}~\bibnamefont
  {Wang}}\ and\ \bibinfo {author} {\bibfnamefont {G.}~\bibnamefont {Cao}},\
  }\href {https://doi.org/10.1103/PhysRevD.97.034014} {\bibfield  {journal}
  {\bibinfo  {journal} {Phys. Rev. D}\ }\textbf {\bibinfo {volume} {97}},\
  \bibinfo {pages} {034014} (\bibinfo {year} {2018})},\ \Eprint
  {https://arxiv.org/abs/1712.09780} {arXiv:1712.09780 [nucl-th]} \BibitemShut
  {NoStop}%
\bibitem [{\citenamefont {Cao}\ and\ \citenamefont {He}(2019)}]{Cao:2019ctl}%
  \BibitemOpen
  \bibfield  {author} {\bibinfo {author} {\bibfnamefont {G.}~\bibnamefont
  {Cao}}\ and\ \bibinfo {author} {\bibfnamefont {L.}~\bibnamefont {He}},\
  }\href {https://doi.org/10.1103/PhysRevD.100.094015} {\bibfield  {journal}
  {\bibinfo  {journal} {Phys. Rev. D}\ }\textbf {\bibinfo {volume} {100}},\
  \bibinfo {pages} {094015} (\bibinfo {year} {2019})},\ \Eprint
  {https://arxiv.org/abs/1910.02728} {arXiv:1910.02728 [nucl-th]} \BibitemShut
  {NoStop}%
\bibitem [{\citenamefont {Chen}\ \emph {et~al.}(2019)\citenamefont {Chen},
  \citenamefont {Huang},\ and\ \citenamefont {Mameda}}]{Chen:2019tcp}%
  \BibitemOpen
  \bibfield  {author} {\bibinfo {author} {\bibfnamefont {H.-L.}\ \bibnamefont
  {Chen}}, \bibinfo {author} {\bibfnamefont {X.-G.}\ \bibnamefont {Huang}},\
  and\ \bibinfo {author} {\bibfnamefont {K.}~\bibnamefont {Mameda}},\
  }\href@noop {} {\  (\bibinfo {year} {2019})},\ \Eprint
  {https://arxiv.org/abs/1910.02700} {arXiv:1910.02700 [nucl-th]} \BibitemShut
  {NoStop}%
\bibitem [{\citenamefont {Chen}\ \emph {et~al.}(2016)\citenamefont {Chen},
  \citenamefont {Fukushima}, \citenamefont {Huang},\ and\ \citenamefont
  {Mameda}}]{Chen:2015hfc}%
  \BibitemOpen
  \bibfield  {author} {\bibinfo {author} {\bibfnamefont {H.-L.}\ \bibnamefont
  {Chen}}, \bibinfo {author} {\bibfnamefont {K.}~\bibnamefont {Fukushima}},
  \bibinfo {author} {\bibfnamefont {X.-G.}\ \bibnamefont {Huang}},\ and\
  \bibinfo {author} {\bibfnamefont {K.}~\bibnamefont {Mameda}},\ }\href
  {https://doi.org/10.1103/PhysRevD.93.104052} {\bibfield  {journal} {\bibinfo
  {journal} {Phys. Rev. D}\ }\textbf {\bibinfo {volume} {93}},\ \bibinfo
  {pages} {104052} (\bibinfo {year} {2016})},\ \Eprint
  {https://arxiv.org/abs/1512.08974} {arXiv:1512.08974 [hep-ph]} \BibitemShut
  {NoStop}%
\bibitem [{\citenamefont {Jiang}\ and\ \citenamefont
  {Liao}(2016)}]{Jiang:2016wvv}%
  \BibitemOpen
  \bibfield  {author} {\bibinfo {author} {\bibfnamefont {Y.}~\bibnamefont
  {Jiang}}\ and\ \bibinfo {author} {\bibfnamefont {J.}~\bibnamefont {Liao}},\
  }\href {https://doi.org/10.1103/PhysRevLett.117.192302} {\bibfield  {journal}
  {\bibinfo  {journal} {Phys. Rev. Lett.}\ }\textbf {\bibinfo {volume} {117}},\
  \bibinfo {pages} {192302} (\bibinfo {year} {2016})},\ \Eprint
  {https://arxiv.org/abs/1606.03808} {arXiv:1606.03808 [hep-ph]} \BibitemShut
  {NoStop}%
\bibitem [{\citenamefont {Bliokh}\ \emph {et~al.}(2011)\citenamefont {Bliokh},
  \citenamefont {Dennis},\ and\ \citenamefont {Nori}}]{PhysRevLett.107.174802}%
  \BibitemOpen
  \bibfield  {author} {\bibinfo {author} {\bibfnamefont {K.~Y.}\ \bibnamefont
  {Bliokh}}, \bibinfo {author} {\bibfnamefont {M.~R.}\ \bibnamefont {Dennis}},\
  and\ \bibinfo {author} {\bibfnamefont {F.}~\bibnamefont {Nori}},\ }\href
  {https://doi.org/10.1103/PhysRevLett.107.174802} {\bibfield  {journal}
  {\bibinfo  {journal} {Phys. Rev. Lett.}\ }\textbf {\bibinfo {volume} {107}},\
  \bibinfo {pages} {174802} (\bibinfo {year} {2011})}\BibitemShut {NoStop}%
\bibitem [{\citenamefont {Hayrapetyan}\ \emph {et~al.}(2014)\citenamefont
  {Hayrapetyan}, \citenamefont {Matula}, \citenamefont {Aiello}, \citenamefont
  {Surzhykov},\ and\ \citenamefont {Fritzsche}}]{PhysRevLett.112.134801}%
  \BibitemOpen
  \bibfield  {author} {\bibinfo {author} {\bibfnamefont {A.~G.}\ \bibnamefont
  {Hayrapetyan}}, \bibinfo {author} {\bibfnamefont {O.}~\bibnamefont {Matula}},
  \bibinfo {author} {\bibfnamefont {A.}~\bibnamefont {Aiello}}, \bibinfo
  {author} {\bibfnamefont {A.}~\bibnamefont {Surzhykov}},\ and\ \bibinfo
  {author} {\bibfnamefont {S.}~\bibnamefont {Fritzsche}},\ }\href
  {https://doi.org/10.1103/PhysRevLett.112.134801} {\bibfield  {journal}
  {\bibinfo  {journal} {Phys. Rev. Lett.}\ }\textbf {\bibinfo {volume} {112}},\
  \bibinfo {pages} {134801} (\bibinfo {year} {2014})}\BibitemShut {NoStop}%
\bibitem [{\citenamefont {Bialynicki-Birula}\ and\ \citenamefont
  {Bialynicka-Birula}(2017)}]{Bialynicki-Birula:2017moy}%
  \BibitemOpen
  \bibfield  {author} {\bibinfo {author} {\bibfnamefont {I.}~\bibnamefont
  {Bialynicki-Birula}}\ and\ \bibinfo {author} {\bibfnamefont {Z.}~\bibnamefont
  {Bialynicka-Birula}},\ }\href
  {https://doi.org/10.1103/PhysRevLett.118.114801} {\bibfield  {journal}
  {\bibinfo  {journal} {Phys. Rev. Lett.}\ }\textbf {\bibinfo {volume} {118}},\
  \bibinfo {pages} {114801} (\bibinfo {year} {2017})},\ \Eprint
  {https://arxiv.org/abs/1611.04445} {arXiv:1611.04445 [quant-ph]} \BibitemShut
  {NoStop}%
\bibitem [{\citenamefont {van Kruining}\ \emph {et~al.}(2017)\citenamefont {van
  Kruining}, \citenamefont {Hayrapetyan},\ and\ \citenamefont
  {G{\"o}tte}}]{vanKruining:2017anw}%
  \BibitemOpen
  \bibfield  {author} {\bibinfo {author} {\bibfnamefont {K.}~\bibnamefont {van
  Kruining}}, \bibinfo {author} {\bibfnamefont {A.~G.}\ \bibnamefont
  {Hayrapetyan}},\ and\ \bibinfo {author} {\bibfnamefont {J.~B.}\ \bibnamefont
  {G{\"o}tte}},\ }\href {https://doi.org/10.1103/PhysRevLett.119.030401}
  {\bibfield  {journal} {\bibinfo  {journal} {Phys.\ Rev.\ Lett.}\ }\textbf
  {\bibinfo {volume} {119}},\ \bibinfo {pages} {030401} (\bibinfo {year}
  {2017})},\ \Eprint {https://arxiv.org/abs/1702.05271} {arXiv:1702.05271
  [quant-ph]} \BibitemShut {NoStop}%
\bibitem [{\citenamefont {Rajabi}\ and\ \citenamefont
  {Berakdar}(2017)}]{PhysRevA.95.063812}%
  \BibitemOpen
  \bibfield  {author} {\bibinfo {author} {\bibfnamefont {A.}~\bibnamefont
  {Rajabi}}\ and\ \bibinfo {author} {\bibfnamefont {J.}~\bibnamefont
  {Berakdar}},\ }\href {https://doi.org/10.1103/PhysRevA.95.063812} {\bibfield
  {journal} {\bibinfo  {journal} {Phys. Rev. A}\ }\textbf {\bibinfo {volume}
  {95}},\ \bibinfo {pages} {063812} (\bibinfo {year} {2017})}\BibitemShut
  {NoStop}%
\bibitem [{\citenamefont {Silenko}\ \emph {et~al.}(2018)\citenamefont
  {Silenko}, \citenamefont {Zhang},\ and\ \citenamefont
  {Zou}}]{PhysRevLett.121.043202}%
  \BibitemOpen
  \bibfield  {author} {\bibinfo {author} {\bibfnamefont {A.~J.}\ \bibnamefont
  {Silenko}}, \bibinfo {author} {\bibfnamefont {P.}~\bibnamefont {Zhang}},\
  and\ \bibinfo {author} {\bibfnamefont {L.}~\bibnamefont {Zou}},\ }\href
  {https://doi.org/10.1103/PhysRevLett.121.043202} {\bibfield  {journal}
  {\bibinfo  {journal} {Phys. Rev. Lett.}\ }\textbf {\bibinfo {volume} {121}},\
  \bibinfo {pages} {043202} (\bibinfo {year} {2018})}\BibitemShut {NoStop}%
\bibitem [{\citenamefont {Silenko}\ \emph {et~al.}(2019)\citenamefont
  {Silenko}, \citenamefont {Zhang},\ and\ \citenamefont
  {Zou}}]{PhysRevLett.122.063201}%
  \BibitemOpen
  \bibfield  {author} {\bibinfo {author} {\bibfnamefont {A.~J.}\ \bibnamefont
  {Silenko}}, \bibinfo {author} {\bibfnamefont {P.}~\bibnamefont {Zhang}},\
  and\ \bibinfo {author} {\bibfnamefont {L.}~\bibnamefont {Zou}},\ }\href
  {https://doi.org/10.1103/PhysRevLett.122.063201} {\bibfield  {journal}
  {\bibinfo  {journal} {Phys. Rev. Lett.}\ }\textbf {\bibinfo {volume} {122}},\
  \bibinfo {pages} {063201} (\bibinfo {year} {2019})}\BibitemShut {NoStop}%
\bibitem [{\citenamefont {Ivanov}\ \emph {et~al.}(2020)\citenamefont {Ivanov},
  \citenamefont {Korchagin}, \citenamefont {Pimikov},\ and\ \citenamefont
  {Zhang}}]{Ivanov:2019vxe}%
  \BibitemOpen
  \bibfield  {author} {\bibinfo {author} {\bibfnamefont {I.~P.}\ \bibnamefont
  {Ivanov}}, \bibinfo {author} {\bibfnamefont {N.}~\bibnamefont {Korchagin}},
  \bibinfo {author} {\bibfnamefont {A.}~\bibnamefont {Pimikov}},\ and\ \bibinfo
  {author} {\bibfnamefont {P.}~\bibnamefont {Zhang}},\ }\href
  {https://doi.org/10.1103/PhysRevLett.124.192001} {\bibfield  {journal}
  {\bibinfo  {journal} {Phys. Rev. Lett.}\ }\textbf {\bibinfo {volume} {124}},\
  \bibinfo {pages} {192001} (\bibinfo {year} {2020})},\ \Eprint
  {https://arxiv.org/abs/1911.08423} {arXiv:1911.08423 [hep-ph]} \BibitemShut
  {NoStop}%
\bibitem [{\citenamefont {Bliokh}\ \emph {et~al.}(2007)\citenamefont {Bliokh},
  \citenamefont {Bliokh}, \citenamefont {Savel'ev},\ and\ \citenamefont
  {Nori}}]{Bliokh:2007ec}%
  \BibitemOpen
  \bibfield  {author} {\bibinfo {author} {\bibfnamefont {K.~{\relax Yu}.}\
  \bibnamefont {Bliokh}}, \bibinfo {author} {\bibfnamefont {Y.~P.}\
  \bibnamefont {Bliokh}}, \bibinfo {author} {\bibfnamefont {S.}~\bibnamefont
  {Savel'ev}},\ and\ \bibinfo {author} {\bibfnamefont {F.}~\bibnamefont
  {Nori}},\ }\href {https://doi.org/10.1103/PhysRevLett.99.190404} {\bibfield
  {journal} {\bibinfo  {journal} {Phys. Rev. Lett.}\ }\textbf {\bibinfo
  {volume} {99}},\ \bibinfo {pages} {190404} (\bibinfo {year} {2007})},\
  \Eprint {https://arxiv.org/abs/0706.2486} {arXiv:0706.2486 [quant-ph]}
  \BibitemShut {NoStop}%
\bibitem [{\citenamefont {Bliokh}\ \emph
  {et~al.}(2017{\natexlab{a}})\citenamefont {Bliokh}, \citenamefont {Dennis},\
  and\ \citenamefont {Nori}}]{Bliokh:2017sdz}%
  \BibitemOpen
  \bibfield  {author} {\bibinfo {author} {\bibfnamefont {K.~Y.}\ \bibnamefont
  {Bliokh}}, \bibinfo {author} {\bibfnamefont {M.~R.}\ \bibnamefont {Dennis}},\
  and\ \bibinfo {author} {\bibfnamefont {F.}~\bibnamefont {Nori}},\ }\href
  {https://doi.org/10.1103/PhysRevA.96.023622} {\bibfield  {journal} {\bibinfo
  {journal} {Phys. Rev. A}\ }\textbf {\bibinfo {volume} {96}},\ \bibinfo
  {pages} {023622} (\bibinfo {year} {2017}{\natexlab{a}})},\ \Eprint
  {https://arxiv.org/abs/1706.01658} {arXiv:1706.01658 [quant-ph]} \BibitemShut
  {NoStop}%
\bibitem [{\citenamefont {Ducharme}\ \emph {et~al.}(2018)\citenamefont
  {Ducharme}, \citenamefont {da~Paz},\ and\ \citenamefont
  {Hayrapetyan}}]{ducharme2018fractional}%
  \BibitemOpen
  \bibfield  {author} {\bibinfo {author} {\bibfnamefont {R.~J.}\ \bibnamefont
  {Ducharme}}, \bibinfo {author} {\bibfnamefont {I.~G.}\ \bibnamefont
  {da~Paz}},\ and\ \bibinfo {author} {\bibfnamefont {A.~G.}\ \bibnamefont
  {Hayrapetyan}},\ }\href@noop {} {\  (\bibinfo {year} {2018})},\ \Eprint
  {https://arxiv.org/abs/1812.04957} {arXiv:1812.04957 [quant-ph]} \BibitemShut
  {NoStop}%
\bibitem [{\citenamefont {Allen}\ \emph {et~al.}(1992)\citenamefont {Allen},
  \citenamefont {Beijersbergen}, \citenamefont {Spreeuw},\ and\ \citenamefont
  {Woerdman}}]{PhysRevA.45.8185}%
  \BibitemOpen
  \bibfield  {author} {\bibinfo {author} {\bibfnamefont {L.}~\bibnamefont
  {Allen}}, \bibinfo {author} {\bibfnamefont {M.~W.}\ \bibnamefont
  {Beijersbergen}}, \bibinfo {author} {\bibfnamefont {R.~J.~C.}\ \bibnamefont
  {Spreeuw}},\ and\ \bibinfo {author} {\bibfnamefont {J.~P.}\ \bibnamefont
  {Woerdman}},\ }\href {https://doi.org/10.1103/PhysRevA.45.8185} {\bibfield
  {journal} {\bibinfo  {journal} {Phys. Rev. A}\ }\textbf {\bibinfo {volume}
  {45}},\ \bibinfo {pages} {8185} (\bibinfo {year} {1992})}\BibitemShut
  {NoStop}%
\bibitem [{\citenamefont {Bliokh}\ \emph {et~al.}(2014)\citenamefont {Bliokh},
  \citenamefont {Dressel},\ and\ \citenamefont {Nori}}]{Bliokh:2014ara}%
  \BibitemOpen
  \bibfield  {author} {\bibinfo {author} {\bibfnamefont {K.~Y.}\ \bibnamefont
  {Bliokh}}, \bibinfo {author} {\bibfnamefont {J.}~\bibnamefont {Dressel}},\
  and\ \bibinfo {author} {\bibfnamefont {F.}~\bibnamefont {Nori}},\ }\href
  {https://doi.org/10.1088/1367-2630/16/9/093037} {\bibfield  {journal}
  {\bibinfo  {journal} {New J. Phys.}\ }\textbf {\bibinfo {volume} {16}},\
  \bibinfo {pages} {093037} (\bibinfo {year} {2014})},\ \Eprint
  {https://arxiv.org/abs/1404.5486} {arXiv:1404.5486 [physics.optics]}
  \BibitemShut {NoStop}%
\bibitem [{\citenamefont {Barnett}(2017)}]{Barnett:2017wrr}%
  \BibitemOpen
  \bibfield  {author} {\bibinfo {author} {\bibfnamefont {S.~M.}\ \bibnamefont
  {Barnett}},\ }\href {https://doi.org/10.1103/PhysRevLett.118.114802}
  {\bibfield  {journal} {\bibinfo  {journal} {Phys. Rev. Lett.}\ }\textbf
  {\bibinfo {volume} {118}},\ \bibinfo {pages} {114802} (\bibinfo {year}
  {2017})}\BibitemShut {NoStop}%
\bibitem [{\citenamefont {Uchida}\ and\ \citenamefont
  {Tonomura}(2010)}]{uchida2010generation}%
  \BibitemOpen
  \bibfield  {author} {\bibinfo {author} {\bibfnamefont {M.}~\bibnamefont
  {Uchida}}\ and\ \bibinfo {author} {\bibfnamefont {A.}~\bibnamefont
  {Tonomura}},\ }\href {https://doi.org/10.1038/nature08904} {\bibfield
  {journal} {\bibinfo  {journal} {Nature}\ }\textbf {\bibinfo {volume} {464}},\
  \bibinfo {pages} {737} (\bibinfo {year} {2010})}\BibitemShut {NoStop}%
\bibitem [{\citenamefont {Verbeeck}\ \emph {et~al.}(2010)\citenamefont
  {Verbeeck}, \citenamefont {Tian},\ and\ \citenamefont
  {Schattschneider}}]{verbeeck2010production}%
  \BibitemOpen
  \bibfield  {author} {\bibinfo {author} {\bibfnamefont {J.}~\bibnamefont
  {Verbeeck}}, \bibinfo {author} {\bibfnamefont {H.}~\bibnamefont {Tian}},\
  and\ \bibinfo {author} {\bibfnamefont {P.}~\bibnamefont {Schattschneider}},\
  }\href {https://doi.org/10.1038/nature09366} {\bibfield  {journal} {\bibinfo
  {journal} {Nature}\ }\textbf {\bibinfo {volume} {467}},\ \bibinfo {pages}
  {301} (\bibinfo {year} {2010})}\BibitemShut {NoStop}%
\bibitem [{\citenamefont {McMorran}\ \emph {et~al.}(2011)\citenamefont
  {McMorran}, \citenamefont {Agrawal}, \citenamefont {Anderson}, \citenamefont
  {Herzing}, \citenamefont {Lezec}, \citenamefont {McClelland},\ and\
  \citenamefont {Unguris}}]{mcmorran2011electron}%
  \BibitemOpen
  \bibfield  {author} {\bibinfo {author} {\bibfnamefont {B.~J.}\ \bibnamefont
  {McMorran}}, \bibinfo {author} {\bibfnamefont {A.}~\bibnamefont {Agrawal}},
  \bibinfo {author} {\bibfnamefont {I.~M.}\ \bibnamefont {Anderson}}, \bibinfo
  {author} {\bibfnamefont {A.~A.}\ \bibnamefont {Herzing}}, \bibinfo {author}
  {\bibfnamefont {H.~J.}\ \bibnamefont {Lezec}}, \bibinfo {author}
  {\bibfnamefont {J.~J.}\ \bibnamefont {McClelland}},\ and\ \bibinfo {author}
  {\bibfnamefont {J.}~\bibnamefont {Unguris}},\ }\href
  {https://doi.org/10.1126/science.1198804} {\bibfield  {journal} {\bibinfo
  {journal} {Science}\ }\textbf {\bibinfo {volume} {331}},\ \bibinfo {pages}
  {192} (\bibinfo {year} {2011})}\BibitemShut {NoStop}%
\bibitem [{\citenamefont {Clark}\ \emph {et~al.}(2015)\citenamefont {Clark},
  \citenamefont {Barankov}, \citenamefont {Huber}, \citenamefont {Arif},
  \citenamefont {Cory},\ and\ \citenamefont {Pushin}}]{clark2015controlling}%
  \BibitemOpen
  \bibfield  {author} {\bibinfo {author} {\bibfnamefont {C.~W.}\ \bibnamefont
  {Clark}}, \bibinfo {author} {\bibfnamefont {R.}~\bibnamefont {Barankov}},
  \bibinfo {author} {\bibfnamefont {M.~G.}\ \bibnamefont {Huber}}, \bibinfo
  {author} {\bibfnamefont {M.}~\bibnamefont {Arif}}, \bibinfo {author}
  {\bibfnamefont {D.~G.}\ \bibnamefont {Cory}},\ and\ \bibinfo {author}
  {\bibfnamefont {D.~A.}\ \bibnamefont {Pushin}},\ }\href
  {https://doi.org/10.1038/nature15265} {\bibfield  {journal} {\bibinfo
  {journal} {Nature}\ }\textbf {\bibinfo {volume} {525}},\ \bibinfo {pages}
  {504} (\bibinfo {year} {2015})}\BibitemShut {NoStop}%
\bibitem [{\citenamefont {Harris}\ \emph {et~al.}(2015)\citenamefont {Harris},
  \citenamefont {Grillo}, \citenamefont {Mafakheri}, \citenamefont {Gazzadi},
  \citenamefont {Frabboni}, \citenamefont {Boyd},\ and\ \citenamefont
  {Karimi}}]{harris2015structured}%
  \BibitemOpen
  \bibfield  {author} {\bibinfo {author} {\bibfnamefont {J.}~\bibnamefont
  {Harris}}, \bibinfo {author} {\bibfnamefont {V.}~\bibnamefont {Grillo}},
  \bibinfo {author} {\bibfnamefont {E.}~\bibnamefont {Mafakheri}}, \bibinfo
  {author} {\bibfnamefont {G.~C.}\ \bibnamefont {Gazzadi}}, \bibinfo {author}
  {\bibfnamefont {S.}~\bibnamefont {Frabboni}}, \bibinfo {author}
  {\bibfnamefont {R.~W.}\ \bibnamefont {Boyd}},\ and\ \bibinfo {author}
  {\bibfnamefont {E.}~\bibnamefont {Karimi}},\ }\href
  {https://doi.org/10.1038/nphys3404} {\bibfield  {journal} {\bibinfo
  {journal} {Nature Physics}\ }\textbf {\bibinfo {volume} {11}},\ \bibinfo
  {pages} {629} (\bibinfo {year} {2015})}\BibitemShut {NoStop}%
\bibitem [{\citenamefont {Grillo}\ \emph {et~al.}(2015)\citenamefont {Grillo},
  \citenamefont {Gazzadi}, \citenamefont {Mafakheri}, \citenamefont {Frabboni},
  \citenamefont {Karimi},\ and\ \citenamefont {Boyd}}]{PhysRevLett.114.034801}%
  \BibitemOpen
  \bibfield  {author} {\bibinfo {author} {\bibfnamefont {V.}~\bibnamefont
  {Grillo}}, \bibinfo {author} {\bibfnamefont {G.~C.}\ \bibnamefont {Gazzadi}},
  \bibinfo {author} {\bibfnamefont {E.}~\bibnamefont {Mafakheri}}, \bibinfo
  {author} {\bibfnamefont {S.}~\bibnamefont {Frabboni}}, \bibinfo {author}
  {\bibfnamefont {E.}~\bibnamefont {Karimi}},\ and\ \bibinfo {author}
  {\bibfnamefont {R.~W.}\ \bibnamefont {Boyd}},\ }\href
  {https://doi.org/10.1103/PhysRevLett.114.034801} {\bibfield  {journal}
  {\bibinfo  {journal} {Phys. Rev. Lett.}\ }\textbf {\bibinfo {volume} {114}},\
  \bibinfo {pages} {034801} (\bibinfo {year} {2015})}\BibitemShut {NoStop}%
\bibitem [{\citenamefont {Bliokh}\ \emph
  {et~al.}(2017{\natexlab{b}})\citenamefont {Bliokh} \emph
  {et~al.}}]{Bliokh:2017uvr}%
  \BibitemOpen
  \bibfield  {author} {\bibinfo {author} {\bibfnamefont {K.~Y.}\ \bibnamefont
  {Bliokh}} \emph {et~al.},\ }\href
  {https://doi.org/10.1016/j.physrep.2017.05.006} {\bibfield  {journal}
  {\bibinfo  {journal} {Phys. Rept.}\ }\textbf {\bibinfo {volume} {690}},\
  \bibinfo {pages} {1} (\bibinfo {year} {2017}{\natexlab{b}})},\ \Eprint
  {https://arxiv.org/abs/1703.06879} {arXiv:1703.06879 [quant-ph]} \BibitemShut
  {NoStop}%
\bibitem [{\citenamefont {Mafakheri}\ \emph {et~al.}(2017)\citenamefont
  {Mafakheri}, \citenamefont {Tavabi}, \citenamefont {Lu}, \citenamefont
  {Balboni}, \citenamefont {Venturi}, \citenamefont {Menozzi}, \citenamefont
  {Gazzadi}, \citenamefont {Frabboni}, \citenamefont {Sit}, \citenamefont
  {Dunin-Borkowski}, \citenamefont {Karimi},\ and\ \citenamefont
  {Grillo}}]{doi:10.1063/1.4977879}%
  \BibitemOpen
  \bibfield  {author} {\bibinfo {author} {\bibfnamefont {E.}~\bibnamefont
  {Mafakheri}}, \bibinfo {author} {\bibfnamefont {A.~H.}\ \bibnamefont
  {Tavabi}}, \bibinfo {author} {\bibfnamefont {P.-H.}\ \bibnamefont {Lu}},
  \bibinfo {author} {\bibfnamefont {R.}~\bibnamefont {Balboni}}, \bibinfo
  {author} {\bibfnamefont {F.}~\bibnamefont {Venturi}}, \bibinfo {author}
  {\bibfnamefont {C.}~\bibnamefont {Menozzi}}, \bibinfo {author} {\bibfnamefont
  {G.~C.}\ \bibnamefont {Gazzadi}}, \bibinfo {author} {\bibfnamefont
  {S.}~\bibnamefont {Frabboni}}, \bibinfo {author} {\bibfnamefont
  {A.}~\bibnamefont {Sit}}, \bibinfo {author} {\bibfnamefont {R.~E.}\
  \bibnamefont {Dunin-Borkowski}}, \bibinfo {author} {\bibfnamefont
  {E.}~\bibnamefont {Karimi}},\ and\ \bibinfo {author} {\bibfnamefont
  {V.}~\bibnamefont {Grillo}},\ }\href {https://doi.org/10.1063/1.4977879}
  {\bibfield  {journal} {\bibinfo  {journal} {Applied Physics Letters}\
  }\textbf {\bibinfo {volume} {110}},\ \bibinfo {pages} {093113} (\bibinfo
  {year} {2017})}\BibitemShut {NoStop}%
\bibitem [{\citenamefont {Wang}\ \emph
  {et~al.}(2019{\natexlab{b}})\citenamefont {Wang}, \citenamefont {Jiang},
  \citenamefont {He},\ and\ \citenamefont {Zhuang}}]{Wang:2018zrn}%
  \BibitemOpen
  \bibfield  {author} {\bibinfo {author} {\bibfnamefont {L.}~\bibnamefont
  {Wang}}, \bibinfo {author} {\bibfnamefont {Y.}~\bibnamefont {Jiang}},
  \bibinfo {author} {\bibfnamefont {L.}~\bibnamefont {He}},\ and\ \bibinfo
  {author} {\bibfnamefont {P.}~\bibnamefont {Zhuang}},\ }\href
  {https://doi.org/10.1103/PhysRevC.100.034902} {\bibfield  {journal} {\bibinfo
   {journal} {Phys.\ Rev.\ C}\ }\textbf {\bibinfo {volume} {100}},\ \bibinfo
  {pages} {034902} (\bibinfo {year} {2019}{\natexlab{b}})},\ \Eprint
  {https://arxiv.org/abs/1901.00804} {arXiv:1901.00804 [nucl-th]} \BibitemShut
  {NoStop}%
\bibitem [{\citenamefont {Wang}\ \emph
  {et~al.}(2019{\natexlab{c}})\citenamefont {Wang}, \citenamefont {Jiang},
  \citenamefont {He},\ and\ \citenamefont {Zhuang}}]{Wang:2019nhd}%
  \BibitemOpen
  \bibfield  {author} {\bibinfo {author} {\bibfnamefont {L.}~\bibnamefont
  {Wang}}, \bibinfo {author} {\bibfnamefont {Y.}~\bibnamefont {Jiang}},
  \bibinfo {author} {\bibfnamefont {L.}~\bibnamefont {He}},\ and\ \bibinfo
  {author} {\bibfnamefont {P.}~\bibnamefont {Zhuang}},\ }\href
  {https://doi.org/10.1103/PhysRevD.100.114009} {\bibfield  {journal} {\bibinfo
   {journal} {Phys.\ Rev.\ D}\ }\textbf {\bibinfo {volume} {100}},\ \bibinfo
  {pages} {114009} (\bibinfo {year} {2019}{\natexlab{c}})},\ \Eprint
  {https://arxiv.org/abs/1901.04697} {arXiv:1901.04697 [nucl-th]} \BibitemShut
  {NoStop}%
\bibitem [{\citenamefont {Sheng}\ \emph {et~al.}(2018)\citenamefont {Sheng},
  \citenamefont {Rischke}, \citenamefont {Vasak},\ and\ \citenamefont
  {Wang}}]{Sheng:2017lfu}%
  \BibitemOpen
  \bibfield  {author} {\bibinfo {author} {\bibfnamefont {X.-l.}\ \bibnamefont
  {Sheng}}, \bibinfo {author} {\bibfnamefont {D.~H.}\ \bibnamefont {Rischke}},
  \bibinfo {author} {\bibfnamefont {D.}~\bibnamefont {Vasak}},\ and\ \bibinfo
  {author} {\bibfnamefont {Q.}~\bibnamefont {Wang}},\ }\href
  {https://doi.org/10.1140/epja/i2018-12414-9} {\bibfield  {journal} {\bibinfo
  {journal} {Eur. Phys. J. A}\ }\textbf {\bibinfo {volume} {54}},\ \bibinfo
  {pages} {21} (\bibinfo {year} {2018})},\ \Eprint
  {https://arxiv.org/abs/1707.01388} {arXiv:1707.01388 [hep-ph]} \BibitemShut
  {NoStop}%
\bibitem [{\citenamefont {Dong}\ \emph {et~al.}(2020)\citenamefont {Dong},
  \citenamefont {Fang}, \citenamefont {Hou},\ and\ \citenamefont
  {She}}]{Dong:2020zci}%
  \BibitemOpen
  \bibfield  {author} {\bibinfo {author} {\bibfnamefont {R.-D.}\ \bibnamefont
  {Dong}}, \bibinfo {author} {\bibfnamefont {R.-H.}\ \bibnamefont {Fang}},
  \bibinfo {author} {\bibfnamefont {D.-F.}\ \bibnamefont {Hou}},\ and\ \bibinfo
  {author} {\bibfnamefont {D.}~\bibnamefont {She}},\ }\href
  {https://doi.org/10.1088/1674-1137/44/7/074106} {\bibfield  {journal}
  {\bibinfo  {journal} {Chin. Phys. C}\ }\textbf {\bibinfo {volume} {44}},\
  \bibinfo {pages} {074106} (\bibinfo {year} {2020})},\ \Eprint
  {https://arxiv.org/abs/2001.05801} {arXiv:2001.05801 [hep-th]} \BibitemShut
  {NoStop}%
\bibitem [{Note1()}]{Note1}%
  \BibitemOpen
  \bibinfo {note} {This important observation was pointed out to us by
  Kazuya~Mameda.}\BibitemShut {Stop}%
\bibitem [{\citenamefont {Greenshields}\ \emph {et~al.}(2014)\citenamefont
  {Greenshields}, \citenamefont {Stamps}, \citenamefont {Franke-Arnold},\ and\
  \citenamefont {Barnett}}]{PhysRevLett.113.240404}%
  \BibitemOpen
  \bibfield  {author} {\bibinfo {author} {\bibfnamefont {C.~R.}\ \bibnamefont
  {Greenshields}}, \bibinfo {author} {\bibfnamefont {R.~L.}\ \bibnamefont
  {Stamps}}, \bibinfo {author} {\bibfnamefont {S.}~\bibnamefont
  {Franke-Arnold}},\ and\ \bibinfo {author} {\bibfnamefont {S.~M.}\
  \bibnamefont {Barnett}},\ }\href
  {https://doi.org/10.1103/PhysRevLett.113.240404} {\bibfield  {journal}
  {\bibinfo  {journal} {Phys. Rev. Lett.}\ }\textbf {\bibinfo {volume} {113}},\
  \bibinfo {pages} {240404} (\bibinfo {year} {2014})}\BibitemShut {NoStop}%
\bibitem [{\citenamefont {Manuel}\ and\ \citenamefont
  {Torres-Rincon}(2014)}]{Manuel:2013zaa}%
  \BibitemOpen
  \bibfield  {author} {\bibinfo {author} {\bibfnamefont {C.}~\bibnamefont
  {Manuel}}\ and\ \bibinfo {author} {\bibfnamefont {J.~M.}\ \bibnamefont
  {Torres-Rincon}},\ }\href {https://doi.org/10.1103/PhysRevD.89.096002}
  {\bibfield  {journal} {\bibinfo  {journal} {Phys.\ Rev.\ D}\ }\textbf
  {\bibinfo {volume} {89}},\ \bibinfo {pages} {096002} (\bibinfo {year}
  {2014})},\ \Eprint {https://arxiv.org/abs/1312.1158} {arXiv:1312.1158
  [hep-ph]} \BibitemShut {NoStop}%
\bibitem [{\citenamefont {Chen}\ \emph {et~al.}(2014)\citenamefont {Chen},
  \citenamefont {Son}, \citenamefont {Stephanov}, \citenamefont {Yee},\ and\
  \citenamefont {Yin}}]{Chen:2014cla}%
  \BibitemOpen
  \bibfield  {author} {\bibinfo {author} {\bibfnamefont {J.-Y.}\ \bibnamefont
  {Chen}}, \bibinfo {author} {\bibfnamefont {D.~T.}\ \bibnamefont {Son}},
  \bibinfo {author} {\bibfnamefont {M.~A.}\ \bibnamefont {Stephanov}}, \bibinfo
  {author} {\bibfnamefont {H.-U.}\ \bibnamefont {Yee}},\ and\ \bibinfo {author}
  {\bibfnamefont {Y.}~\bibnamefont {Yin}},\ }\href
  {https://doi.org/10.1103/PhysRevLett.113.182302} {\bibfield  {journal}
  {\bibinfo  {journal} {Phys.\ Rev.\ Lett.}\ }\textbf {\bibinfo {volume}
  {113}},\ \bibinfo {pages} {182302} (\bibinfo {year} {2014})},\ \Eprint
  {https://arxiv.org/abs/1404.5963} {arXiv:1404.5963 [hep-th]} \BibitemShut
  {NoStop}%
\bibitem [{\citenamefont {Chen}\ \emph {et~al.}(2015)\citenamefont {Chen},
  \citenamefont {Son},\ and\ \citenamefont {Stephanov}}]{Chen:2015gta}%
  \BibitemOpen
  \bibfield  {author} {\bibinfo {author} {\bibfnamefont {J.-Y.}\ \bibnamefont
  {Chen}}, \bibinfo {author} {\bibfnamefont {D.~T.}\ \bibnamefont {Son}},\ and\
  \bibinfo {author} {\bibfnamefont {M.~A.}\ \bibnamefont {Stephanov}},\ }\href
  {https://doi.org/10.1103/PhysRevLett.115.021601} {\bibfield  {journal}
  {\bibinfo  {journal} {Phys.\ Rev.\ Lett.}\ }\textbf {\bibinfo {volume}
  {115}},\ \bibinfo {pages} {021601} (\bibinfo {year} {2015})},\ \Eprint
  {https://arxiv.org/abs/1502.06966} {arXiv:1502.06966 [hep-th]} \BibitemShut
  {NoStop}%
\bibitem [{\citenamefont {Berger}(1970)}]{PhysRevB.2.4559}%
  \BibitemOpen
  \bibfield  {author} {\bibinfo {author} {\bibfnamefont {L.}~\bibnamefont
  {Berger}},\ }\href {https://doi.org/10.1103/PhysRevB.2.4559} {\bibfield
  {journal} {\bibinfo  {journal} {Phys. Rev. B}\ }\textbf {\bibinfo {volume}
  {2}},\ \bibinfo {pages} {4559} (\bibinfo {year} {1970})}\BibitemShut
  {NoStop}%
\bibitem [{\citenamefont {Nagaosa}\ \emph {et~al.}(2010)\citenamefont
  {Nagaosa}, \citenamefont {Sinova}, \citenamefont {Onoda}, \citenamefont
  {MacDonald},\ and\ \citenamefont {Ong}}]{RevModPhys.82.1539}%
  \BibitemOpen
  \bibfield  {author} {\bibinfo {author} {\bibfnamefont {N.}~\bibnamefont
  {Nagaosa}}, \bibinfo {author} {\bibfnamefont {J.}~\bibnamefont {Sinova}},
  \bibinfo {author} {\bibfnamefont {S.}~\bibnamefont {Onoda}}, \bibinfo
  {author} {\bibfnamefont {A.~H.}\ \bibnamefont {MacDonald}},\ and\ \bibinfo
  {author} {\bibfnamefont {N.~P.}\ \bibnamefont {Ong}},\ }\href
  {https://doi.org/10.1103/RevModPhys.82.1539} {\bibfield  {journal} {\bibinfo
  {journal} {Rev. Mod. Phys.}\ }\textbf {\bibinfo {volume} {82}},\ \bibinfo
  {pages} {1539} (\bibinfo {year} {2010})}\BibitemShut {NoStop}%
\bibitem [{\citenamefont {Xiao}\ \emph {et~al.}(2010)\citenamefont {Xiao},
  \citenamefont {Chang},\ and\ \citenamefont {Niu}}]{RevModPhys.82.1959}%
  \BibitemOpen
  \bibfield  {author} {\bibinfo {author} {\bibfnamefont {D.}~\bibnamefont
  {Xiao}}, \bibinfo {author} {\bibfnamefont {M.-C.}\ \bibnamefont {Chang}},\
  and\ \bibinfo {author} {\bibfnamefont {Q.}~\bibnamefont {Niu}},\ }\href
  {https://doi.org/10.1103/RevModPhys.82.1959} {\bibfield  {journal} {\bibinfo
  {journal} {Rev. Mod. Phys.}\ }\textbf {\bibinfo {volume} {82}},\ \bibinfo
  {pages} {1959} (\bibinfo {year} {2010})}\BibitemShut {NoStop}%
\bibitem [{\citenamefont {Hidaka}\ \emph {et~al.}(2017)\citenamefont {Hidaka},
  \citenamefont {Pu},\ and\ \citenamefont {Yang}}]{Hidaka:2016yjf}%
  \BibitemOpen
  \bibfield  {author} {\bibinfo {author} {\bibfnamefont {Y.}~\bibnamefont
  {Hidaka}}, \bibinfo {author} {\bibfnamefont {S.}~\bibnamefont {Pu}},\ and\
  \bibinfo {author} {\bibfnamefont {D.-L.}\ \bibnamefont {Yang}},\ }\href
  {https://doi.org/10.1103/PhysRevD.95.091901} {\bibfield  {journal} {\bibinfo
  {journal} {Phys.\ Rev.\ D}\ }\textbf {\bibinfo {volume} {95}},\ \bibinfo
  {pages} {091901} (\bibinfo {year} {2017})},\ \Eprint
  {https://arxiv.org/abs/1612.04630} {arXiv:1612.04630 [hep-th]} \BibitemShut
  {NoStop}%
\bibitem [{\citenamefont {Hidaka}\ \emph {et~al.}(2018)\citenamefont {Hidaka},
  \citenamefont {Pu},\ and\ \citenamefont {Yang}}]{Hidaka:2017auj}%
  \BibitemOpen
  \bibfield  {author} {\bibinfo {author} {\bibfnamefont {Y.}~\bibnamefont
  {Hidaka}}, \bibinfo {author} {\bibfnamefont {S.}~\bibnamefont {Pu}},\ and\
  \bibinfo {author} {\bibfnamefont {D.-L.}\ \bibnamefont {Yang}},\ }\href
  {https://doi.org/10.1103/PhysRevD.97.016004} {\bibfield  {journal} {\bibinfo
  {journal} {Phys.\ Rev.\ D}\ }\textbf {\bibinfo {volume} {97}},\ \bibinfo
  {pages} {016004} (\bibinfo {year} {2018})},\ \Eprint
  {https://arxiv.org/abs/1710.00278} {arXiv:1710.00278 [hep-th]} \BibitemShut
  {NoStop}%
\bibitem [{\citenamefont {Hidaka}\ and\ \citenamefont
  {Yang}(2018)}]{Hidaka:2018ekt}%
  \BibitemOpen
  \bibfield  {author} {\bibinfo {author} {\bibfnamefont {Y.}~\bibnamefont
  {Hidaka}}\ and\ \bibinfo {author} {\bibfnamefont {D.-L.}\ \bibnamefont
  {Yang}},\ }\href {https://doi.org/10.1103/PhysRevD.98.016012} {\bibfield
  {journal} {\bibinfo  {journal} {Phys.\ Rev.\ D}\ }\textbf {\bibinfo {volume}
  {98}},\ \bibinfo {pages} {016012} (\bibinfo {year} {2018})},\ \Eprint
  {https://arxiv.org/abs/1801.08253} {arXiv:1801.08253 [hep-th]} \BibitemShut
  {NoStop}%
\bibitem [{\citenamefont {Huang}\ and\ \citenamefont
  {Sadofyev}(2019)}]{Huang:2018aly}%
  \BibitemOpen
  \bibfield  {author} {\bibinfo {author} {\bibfnamefont {X.-G.}\ \bibnamefont
  {Huang}}\ and\ \bibinfo {author} {\bibfnamefont {A.~V.}\ \bibnamefont
  {Sadofyev}},\ }\href {https://doi.org/10.1007/JHEP03(2019)084} {\bibfield
  {journal} {\bibinfo  {journal} {JHEP}\ }\textbf {\bibinfo {volume} {03}},\
  \bibinfo {pages} {084}},\ \Eprint {https://arxiv.org/abs/1805.08779}
  {arXiv:1805.08779 [hep-th]} \BibitemShut {NoStop}%
\bibitem [{\citenamefont {Fukushima}\ and\ \citenamefont
  {Pu}(2020)}]{Fukushima:2020qta}%
  \BibitemOpen
  \bibfield  {author} {\bibinfo {author} {\bibfnamefont {K.}~\bibnamefont
  {Fukushima}}\ and\ \bibinfo {author} {\bibfnamefont {S.}~\bibnamefont {Pu}},\
  }\href@noop {} {\  (\bibinfo {year} {2020})},\ \Eprint
  {https://arxiv.org/abs/2001.00359} {arXiv:2001.00359 [hep-ph]} \BibitemShut
  {NoStop}%
\end{thebibliography}%

\end{document}